\documentclass[preprint,showpacs,preprintnumbers,amsmath,amssymb]{revtex4}

\usepackage{amsmath}
\usepackage{natbib}
\usepackage{graphicx}
\usepackage{epstopdf}
\usepackage{subfigure}
\usepackage{mathrsfs}
\usepackage{dcolumn}
\usepackage{bm}
\usepackage{color}
\usepackage[section]{placeins}
\begin{document}
\title{Influence of the external electromagnetic field on the properties of the Novikov-Thorne accretion disk in Kerr spacetime}
\author{Shiyang Hu$^{1}$}
\email{husy_arcturus@163.com}
\author{Yuxiang Zuo$^{1}$}
\author{Dan Li$^{1}$}
\author{Chen Deng$^{2}$}
 \affiliation{
1. School of Mathematics and Physics, University of South China, Hengyang, 421001 People's Republic of China \\
2. School of Astronomy and Space Science, Nanjing University, Nanjing, 210023 People's Republic of China}

\begin{abstract}
The Novikov-Thorne accretion disk model is widely employed in astrophysics, yet computing its blackbody spectrum theoretically requires analytical expressions for the orbital parameters---specific energy, angular momentum, and angular velocity---of the constituent timelike particles, a task extremely challenging in non-integrable curved spacetimes. In this work, we numerically obtain these orbital parameters for quasi-Keplerian motion in Kerr spacetime with an asymptotically uniform magnetic field using iterative, finite-difference, and interpolation methods, enabling simulations of the disk's energy flux density, temperature, and blackbody spectra across diverse spin parameters, observational inclinations, and magnetic field strengths. We demonstrate that when the magnetic field aligns with the black hole's angular momentum, the disk's radiation positively correlates with field strength, while spectral analysis for our specific black hole mass and accretion rate reveals a conservative detectable threshold of $1.0638 \times 10^{-9}$ T for ambient magnetic fields. This study not only extends the Novikov-Thorne model to non-integrable axisymmetric spacetimes but also establishes the first direct relationship between external magnetic fields and disk properties, providing critical theoretical support for future magnetic environment studies through disk radiation observations.
\end{abstract}


\maketitle

\section{Introduction}
From the Laser Interferometer Gravitational-Wave Observatory (LIGO) capturing gravitational waves from binary black hole mergers \cite{Abbott et al. (2016)}, to the Event Horizon Telescope (EHT) ``snapping the shutter'' on the supermassive black holes in M87 and our Galactic Center \cite{Akiyama et al. (2019),Akiyama et al. (2022)}, increasingly sophisticated techniques have been employed to study black holes and their extreme environments. In fact, another powerful approach lies in analyzing the radiation spectra from accretion disks surrounding black holes. These spectral signatures encode vital spacetime information, proving invaluable for black hole diagnostics and parameter estimation.

Building upon the angular momentum transport mechanism, Shakura and Sunyaev proposed the standard thin disk model, whose radiative properties depend on parameters such as the viscosity parameter and accretion rate \cite{Shakura and Sunyaev (1973)}. Novikov and Thorne subsequently generalized this model within general relativity by incorporating conservation laws of energy, angular momentum, and rest mass \cite{Novikov and Thorne (1973)}, establishing what is now known as the Novikov-Thorne accretion disk---a geometrically thin, optically thick structure around black holes. This disk's radiation, effectively modeled as blackbody emission, exhibits such high radiative efficiency that thermal energy cannot accumulate, allowing the disk to maintain stable geometric thinness in the black hole's equatorial plane. The Novikov-Thorne model represents more than theoretical progress; it has achieved remarkable success in observational applications \cite{Li et al. (2005)}, particularly through continuum-fitting methods \cite{Gou et al. (2011),Bambi and Barausse (2011),Gou et al. (2014),Bambi et al. (2016),Tripathi et al. (2022),Zhao et al. (2021),Mummery et al. (2024)}. These demonstrated capabilities warrant continued scientific focus on its characteristic features.

Early studies of Novikov-Thorne accretion disks primarily focused on the coevolution of the disk and central compact object. As Thorne demonstrated \cite{Thorne (1974)}, while accreting matter can spin up a black hole, this acceleration ceases when the spin parameter reaches $a = 0.998$ M (where M is the black hole mass) due to a counteracting torque between accretion and disk radiation. This critical spin corresponds to an approximately $30\%$ efficiency in converting accreted mass-energy into radiation. In recent decades, research emphasis has shifted toward probing the fundamental connection between Novikov-Thorne disk properties and black hole spacetime geometry. Scientists now increasingly leverage disk observations to constrain intrinsic black hole characteristics, particularly through spectral signatures that encode spacetime information. In \cite{Torres (2002)}, the authors systematically compared Novikov-Thorne accretion disk properties around boson stars and Schwarzschild black holes, finding nearly identical luminosities in radio and near-infrared bands but slightly stronger ultraviolet emission for boson stars---a discovery that subsequently motivated broader applications of this framework to distinguish black holes from regular black holes \cite{Boshkayev et al. (2024a),Capozziello et al. (2025)}, wormholes \cite{Harko et al. (2008),Harko et al. (2009a),Karimov et al. (2019),Karimov et al. (2020),Bambhaniya et al. (2022)}, naked singularities \cite{Kovacs and Harko (2010),Joshi et al. (2014),Tahelyani et al. (2022)}, and gravastars \cite{Harko et al. (2009b)}. Within modified gravity theories, where black holes may carry additional ``hair'' parameters beyond mass, spin, and charge, significant attention has focused on how these parameters influence disk characteristics---including temperature, luminosity, and energy flux---providing critical tests of the no-hair theorem \cite{Pun et al. (2008a),Pun et al. (2008b),Harko et al. (2009c),Harko et al. (2010),Chen and Jing (2011),Chen and Jing (2012),Perez et al. (2013),Moore and Gair (2015),Perez et al. (2017),Karimov et al. (2018),Heydari-Fard et al. (2020),Zuluaga and Sanchez (2021),Collodel et al. (2021),Boshkayev et al. (2021a),Liu et al. (2021),Heydari-Fard and Sepangi (2021),Heydari-Fard et al. (2021),Kazempour et al. (2022),He et al. (2022),Fathi and Cruz (2023),Cimdiker et al. (2023),Alloqulov et al. (2024),Wu et al. (2024),Heydari-Fard et al. (2024a),Dyadina and Avdeev (2024),Boshkayev et al. (2024b),Liu et al. (2024),Jiang and Wang (2024),Shu and Huang (2025),Uktamov et al. (2025),Faraji (2025),Nozari et al. (2025b),Nozari et al. (2025a),Liu and Huang (2025),Tang et al. (2025),Kumar (2025)}. Researchers have also identified spectral deviations induced by exotic environments such as dark matter or plasma \cite{Boshkayev et al. (2021b),Boshkayev et al. (2022),Narzilloev and Ahmedov (2022),Heydari-Fard et al. (2024b),Kobialko et al. (2025),Capolup et al. (2025)}. Furthermore, numerous black hole image simulations based on the Novikov-Thorne model \cite{Luminet (1979),Zhang et al. (2021),Zhang et al. (2022),Hu et al. (2023),Guo et al. (2023),Huang et al. (2023),Cui et al. (2024),Li (2024),Li and Guo (2025),Yin et al. (2025),Cai et al. (2025a),Cai et al. (2025b)} have not only expanded its applications but also established novel correlations between spacetime parameters and disk observables.

In summary, the bridge connecting black hole parameters with Novikov-Thorne accretion disk spectra has been largely established within both general relativity and modified gravity frameworks. However, the story remains incomplete. In reality, charged plasmas within accretion disks can generate toroidal currents, endowing the black hole spacetime with additional electromagnetic fields \cite{Petterson (1974)}. Similarly, a black hole can also acquire a magnetic field environment when moving in the vicinity of a magnetar \cite{Stuchlik and Kolos (2016)}. Investigating how such electromagnetic fields influence accretion disk spectra is therefore crucial---it enables spectral diagnostics of magnetized black hole environments. Surprisingly, this topic has received scant attention in the scientific community, primarily due to a fundamental computational obstacle: deriving accretion disk spectra requires analytical expressions for the circular orbit parameters (specific energy, specific angular momentum, angular velocity, and the derivatives of the last two quantities) of timelike particles, a task rendered intractable by the spacetime non-integrability induced by external electromagnetic fields. It should be noted that while current research has indeed investigated accretion disk emission features in black hole solutions coupled with nonlinear electromagnetic fields \cite{Abbas et al. (2023),Ditta et al. (2023),Uniyal et al. (2023),Kurmanov et al. (2024)}, such electromagnetic field configurations typically preserve spacetime integrability---making the addressed problems fundamentally distinct from the challenges confronted in our present work. In this paper, we overcome this challenge by employing Newton iteration and finite-difference method to numerically compute disks' orbital parameters, circumventing the need for analytical solutions. This approach enables the first systematic calculations of energy flux density, temperature, and blackbody spectra for Novikov-Thorne accretion disks around Kerr black holes immersed in asymptotically uniform electromagnetic fields across broad parameter spaces, ultimately revealing how electromagnetic environments shape these fundamental observables.

The paper is organized as follows. In Sec. II, we briefly review the Kerr spacetime immersed in an asymptotically uniform electromagnetic field and analyze how system parameters influence the dynamics of timelike particles. Sec. III presents our numerical simulations of the accretion disk's energy flux density, temperature, and blackbody spectrum within the Novikov-Thorne framework, along with their astrophysical implications. We conclude with a summary and discussion in the final section.
\section{Kerr spacetime with an external electromagnetic field}
In Boyer-Lindquist coordinates $x^{\mu}=(t,r,\theta,\varphi)$, the covariant metric tensor $g_{\mu\nu}$ of the Kerr spacetime is given by
\begin{equation}
g_{\mu\nu}=
\begin{pmatrix}\label{1}
-\left(1-\frac{2r}{\Sigma}\right) & 0 & 0 & -\frac{2ar\sin^{2}\theta}{\Sigma} \\
0 & \frac{\Sigma}{\Delta} & 0 & 0 \\
0 & 0 & \Sigma & 0 \\
-\frac{2ar\sin^{2}\theta}{\Sigma} & 0 & 0 & \left(\rho^{2}+\frac{2ra^{2}}{\Sigma}\sin^{2}\theta\right)\sin^{2}\theta
\end{pmatrix},
\end{equation}
where $a$ is the dimensionless spin parameter, and the functions $\Sigma$, $\rho^{2}$, and $\Delta$ are defined as
\begin{equation}\label{2}
\Sigma = r^{2}+a^{2}\cos^{2}\theta, \quad \rho^{2} = r^{2}+a^{2}, \quad \Delta = \rho^{2}-2r.
\end{equation}

We consider a Kerr black hole immersed in an external asymptotically uniform magnetic field described by the Wald potential \cite{Wald (1974)}. The magnetic field is aligned with the black hole's angular momentum, and the electromagnetic four-vector potential takes the form
\begin{equation}\label{3}
A^{\mu}=aB\xi^{\mu}_{(t)} + \frac{B}{2}\xi^{\mu}_{(\varphi)},
\end{equation}
where $B$ represents the magnetic field strength. Since the Kerr spacetime admits two Killing vector fields, $\xi^{\mu}_{(t)} = (1,0,0,0)$ and $\xi^{\mu}_{(\varphi)} = (0,0,0,1)$, the electromagnetic four-vector potential has only two non-vanishing components:
\begin{eqnarray}\label{4}
A_{t} = -aB\left[1+\frac{r}{\Sigma}\left(\sin^{2}\theta-2\right)\right], \\
A_{\varphi} = B\sin^{2}\theta\left[\frac{r^{2}+a^{2}}{2}+\frac{a^{2}r}{\Sigma}\left(\sin^{2}\theta-2\right)\right].
\end{eqnarray}
Hence, for a Kerr black hole with an external magnetic field, the motion of timelike particles is governed by the super-Hamiltonian
\begin{equation}\label{5}
\mathscr{H} = \frac{1}{2}g^{\mu\nu}\left(p_{\mu}-qA_{\mu}\right)\left(p_{\nu}-qA_{\nu}\right),
\end{equation}
where $q$ and $p_{\mu}$ denote the particle's charge and conjugate momentum, respectively. Here $g^{\mu\nu}$ is the contravariant metric tensor
\begin{equation}
g^{\mu\nu}=
\begin{pmatrix}\label{6}
-\frac{\Xi}{\Delta\Sigma} & 0 & 0 & -\frac{2ar}{\Delta\Sigma} \\
0 & \frac{\Delta}{\Sigma} & 0 & 0 \\
0 & 0 & \frac{1}{\Sigma} & 0 \\
-\frac{2ar}{\Delta\Sigma} & 0 & 0 & \frac{\Sigma-2r}{\Delta\Sigma\sin^{2}\theta}
\end{pmatrix},
\end{equation}
with the expression $\Xi$ defined as 
\begin{equation}\label{7}
\Xi = \left(r^{2}+a^{2}\right)^{2}-\Delta a^{2}\sin^{2}\theta.
\end{equation}

Next, we define the effective potential $V_{\textrm{eff}}$, a key quantity for determining the parameters of circular particle orbits---essential for studying Novikov-Thorne accretion disks. Physically, $V_{\textrm{eff}}$ is equivalent to the particle's specific energy $E$, which relates to the electromagnetic potential, conjugate momentum, and four-velocity through
\begin{equation}\label{8}
V_{\textrm{eff}} = E = -p_{t} = -\left(g_{tt}\dot{t} + g_{t\varphi}\dot{\varphi} + qA_{t}\right),
\end{equation}
where the overdot denotes derivatives with respect to proper time $\tau$. The effective potential can be derived by combining the super-Hamiltonian with the timelike particle constraint $\mathscr{H}=-1/2$, yielding
\begin{equation}\label{9}
V_{\textrm{eff}} = -\frac{-f_{2}+\sqrt{f^{2}_{2}-4f_{1}f_{3}}}{2f_{1}}.
\end{equation}
Here, the coefficient functions $f_{1}$, $f_{2}$, and $f_{3}$ are defined as
\begin{eqnarray}\label{10}
f_{1} = g^{tt}, \\
f_{2} = 2g^{t\varphi}p_{\varphi}-2g^{t\varphi}qA_{\varphi}-2g^{tt}qA_{t}, \\
f_{3} = g^{tt}q^{2}A^{2}_{t}+g^{\varphi\varphi}\left(p_{\varphi}-qA_{\varphi}\right)^{2}-2g^{t\varphi}qA_{t}p_{\varphi}+2g^{t\varphi}q^{2}A_{t}A_{\varphi}+1.
\end{eqnarray}
Clearly, the effective potential is a function of the particle's radial coordinate $r$, polar angle $\theta$, and specific angular momentum $p_{\varphi}$. Analogous to Eq. \eqref{8}, the quantity $p_{\varphi}$ is given by
\begin{equation}\label{11}
p_{\varphi} = g_{t\varphi}\dot{t}+g_{\varphi\varphi}\dot{\varphi}+qA_{\varphi}.
\end{equation}
We note that the charge $q$ and magnetic field strength $B$ always appear in coupled form. To streamline subsequent calculations, we therefore introduce a new magnetic parameter $\beta=qB$.
\begin{figure*}
\center{
\includegraphics[width=4.5cm]{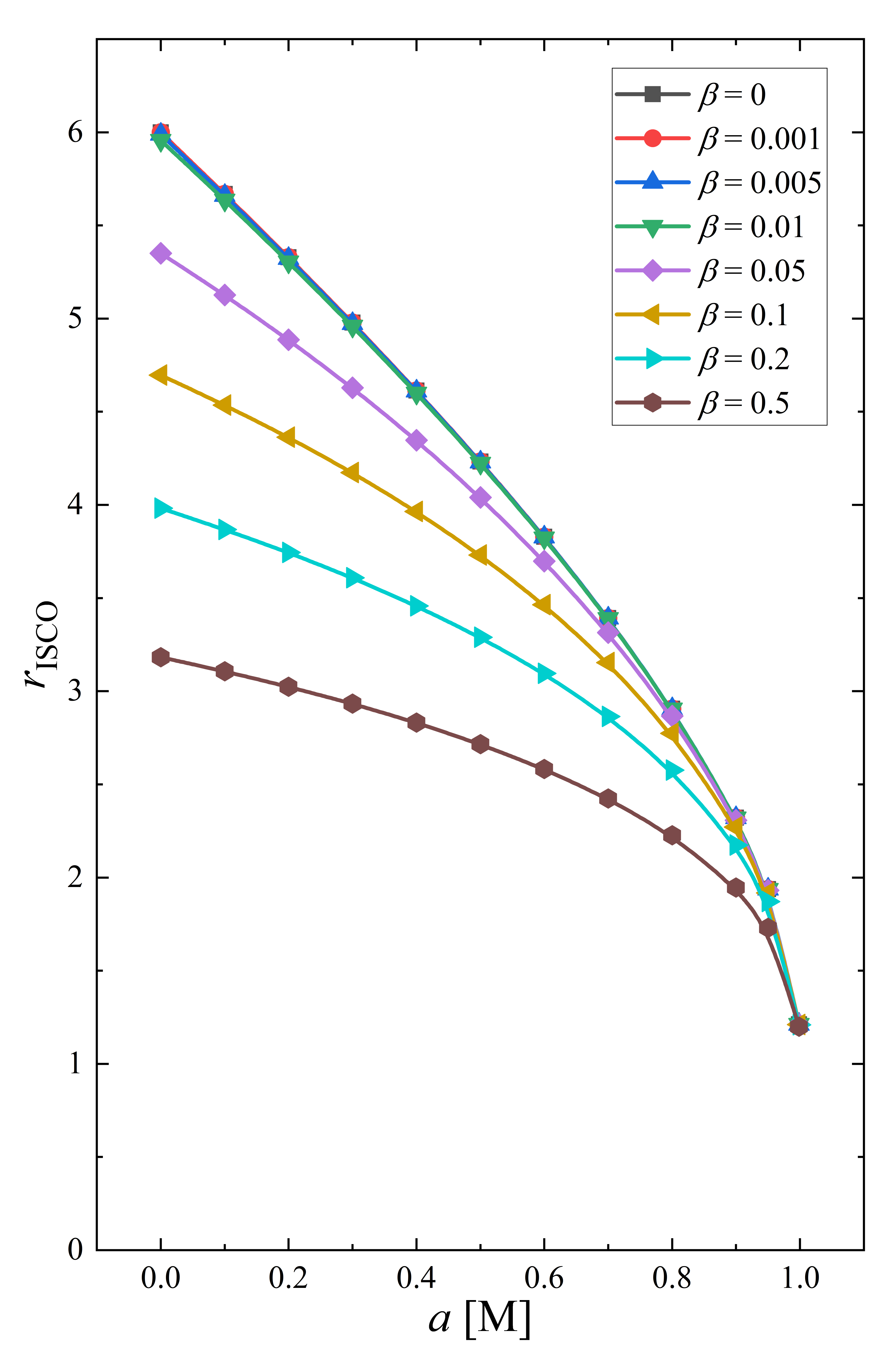}
\caption{(colour online) Dependence of ISCO radius on spin parameter $a$ for varying magnetic parameters $\beta$. The ISCO radius decreases with both increasing spin and magnetic field strength, while magnetic effects become negligible when $\beta \leq 0.01$. Notably, stronger magnetic fields suppress the spin's influence on ISCO location.}} \label{fig1}
\end{figure*}

In the Novikov-Thorne accretion disk model, the high radiative cooling efficiency prevents vertical expansion of the disk due to overheating. Consequently, the disk can be considered to remain strictly confined to the equatorial plane, maintaining $\theta=\pi/2$ throughout evolution. The fundamental constituents of the disk are timelike particles moving along circular orbits at different radii, each corresponding to a local minimum of the effective potential. Notably, the innermost stable circular orbit (ISCO) serves as the disk's inner boundary, whose radius is determined by the simultaneous conditions $\partial V_{\textrm{eff}}/\partial r = \partial^{2} V_{\textrm{eff}}/\partial r^{2} = 0$. Fig. 1 illustrates the dependence of the ISCO radius on both the spin parameter and magnetic field strength, clearly demonstrating an inverse correlation between $r_{\textrm{ISCO}}$ and these two parameters. Specifically, when $\beta$ is relatively small ($\beta \leq 0.01$), its suppressive effect on the ISCO radius is nearly negligible. As the magnetic field strength gradually increases beyond this range, the reduction in $r_{\textrm{ISCO}}$ becomes significantly more pronounced. In contrast, the ISCO radius exhibits noticeable variation across the entire range of spin parameter values. Furthermore, when the spin parameter becomes sufficiently large (i.e., $a > 0.95$), the magnetic field's influence on $r_{\textrm{ISCO}}$ becomes barely noticeable. The results presented in Fig. 1 also confirm the existence of degeneracy between magnetic field strength and spin parameter.

\begin{figure*}
\center{
\includegraphics[width=5cm]{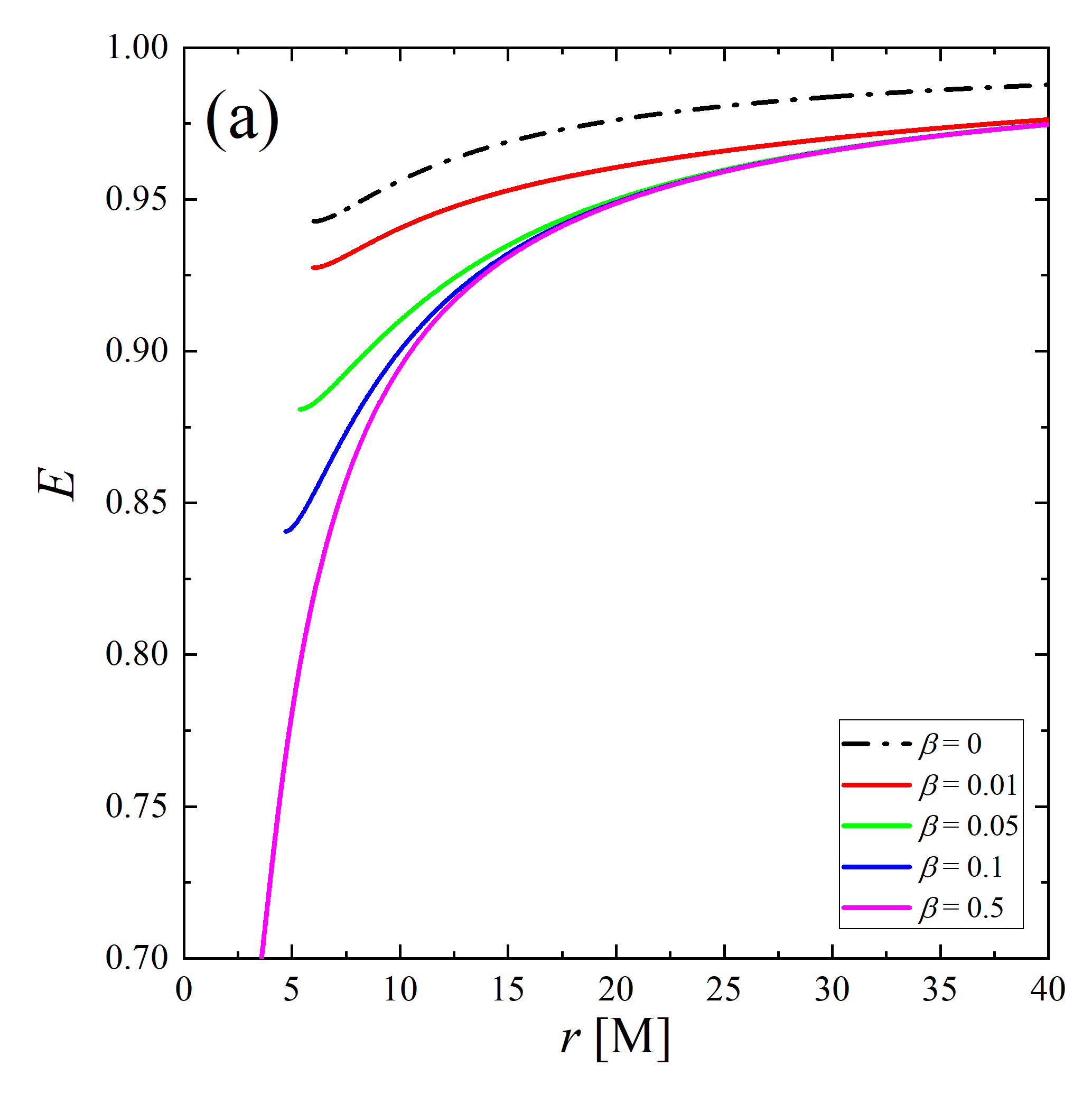}
\includegraphics[width=5cm]{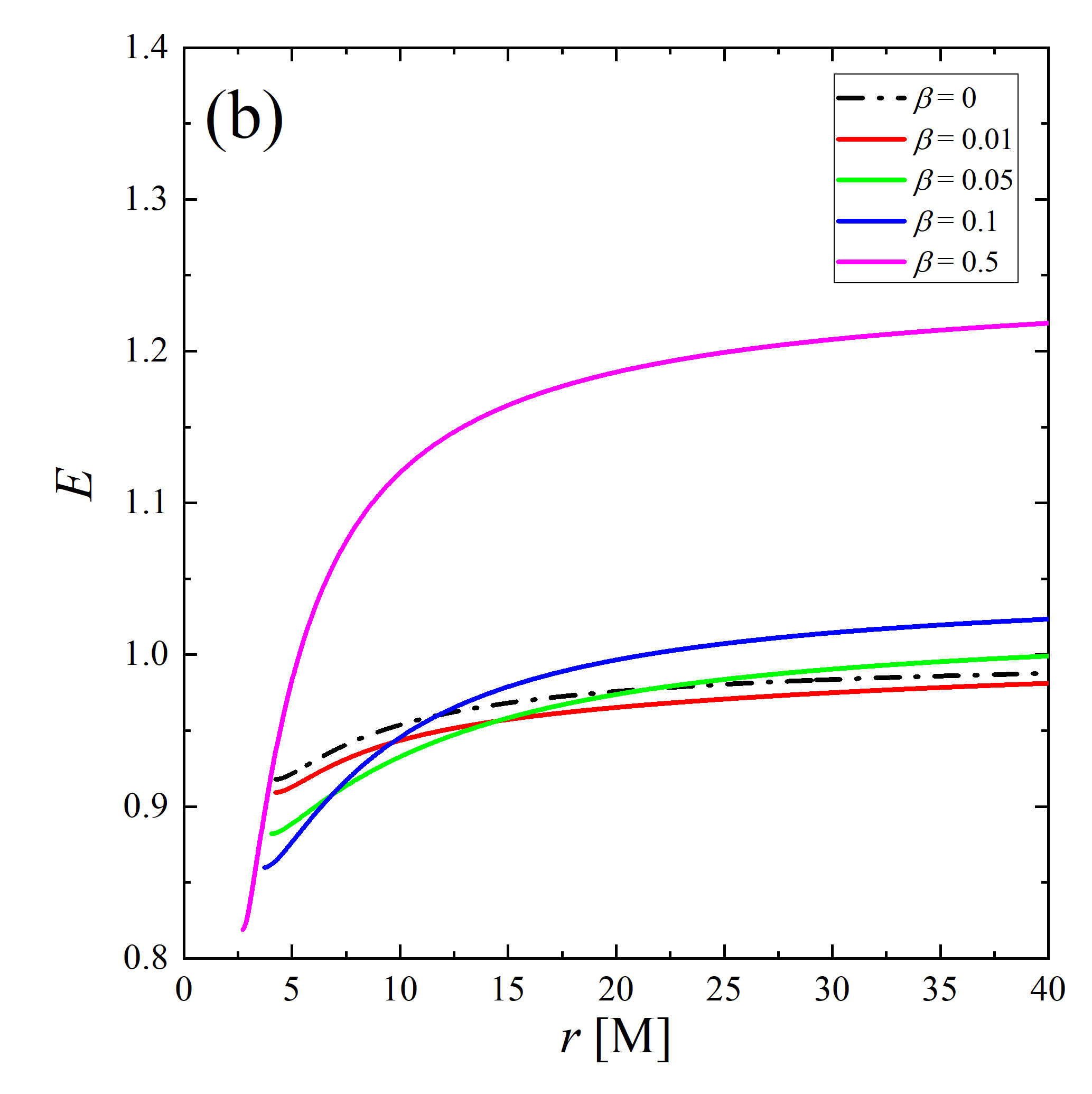}
\includegraphics[width=5cm]{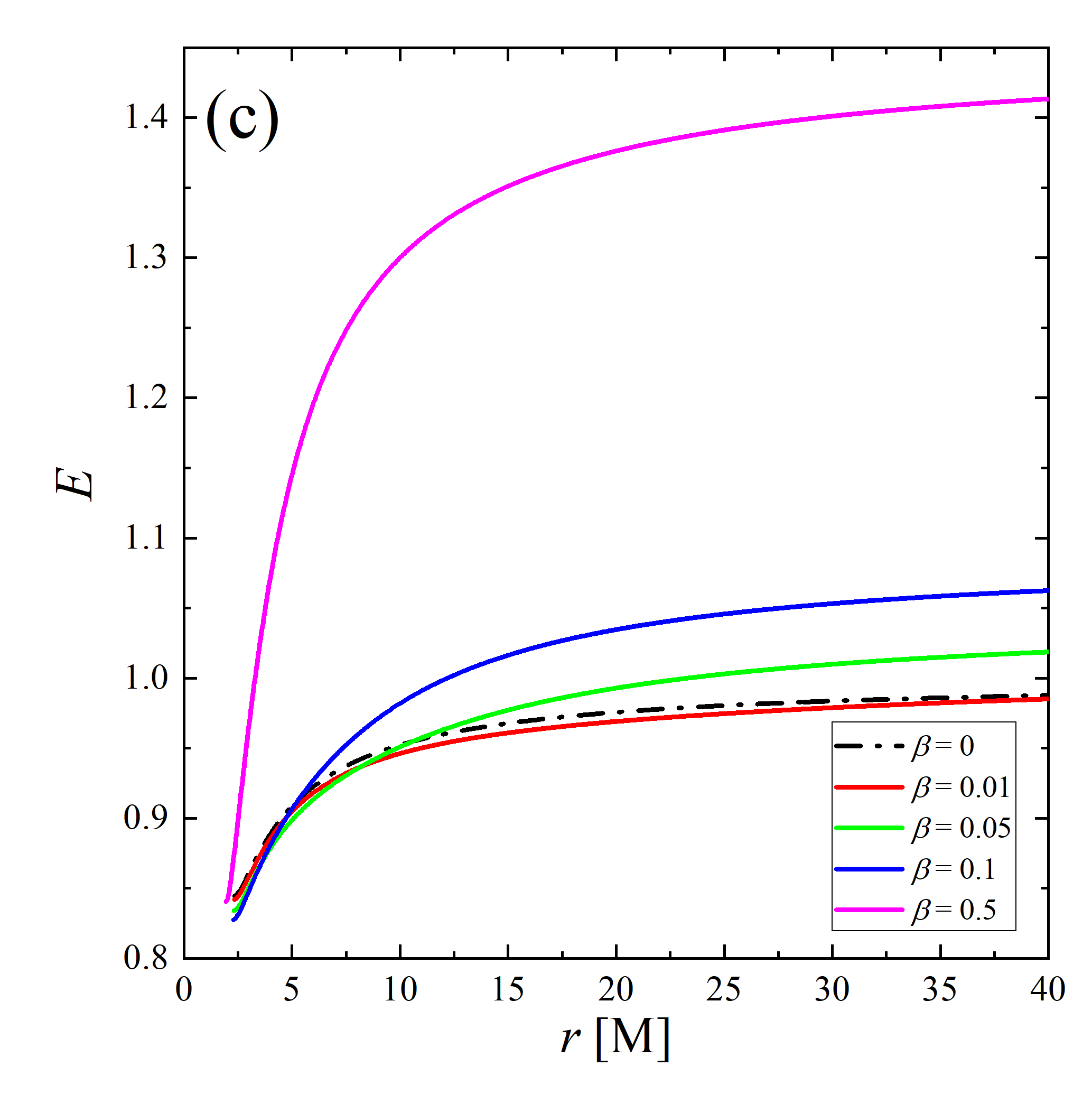}
\includegraphics[width=5cm]{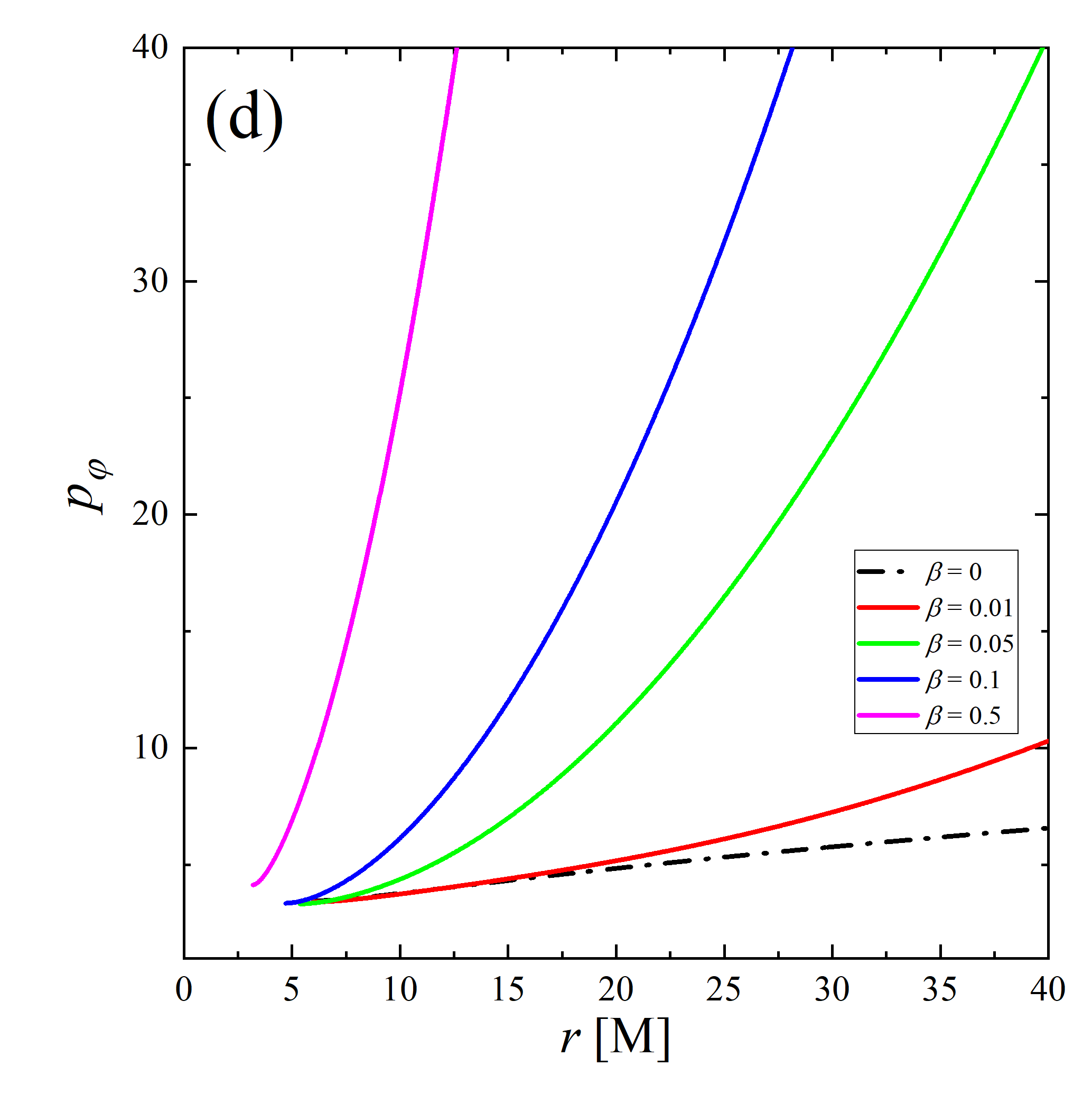}
\includegraphics[width=5cm]{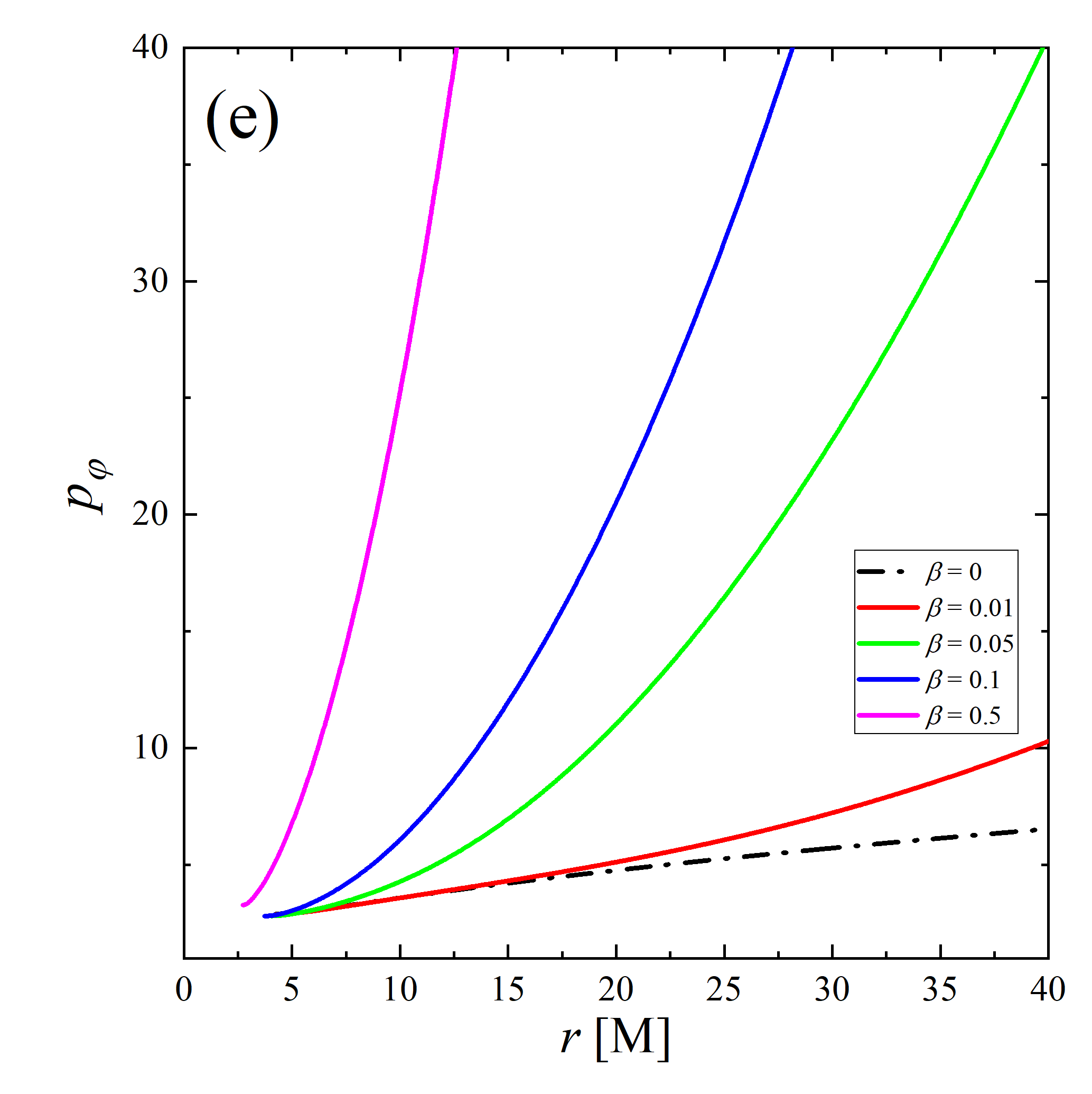}
\includegraphics[width=5cm]{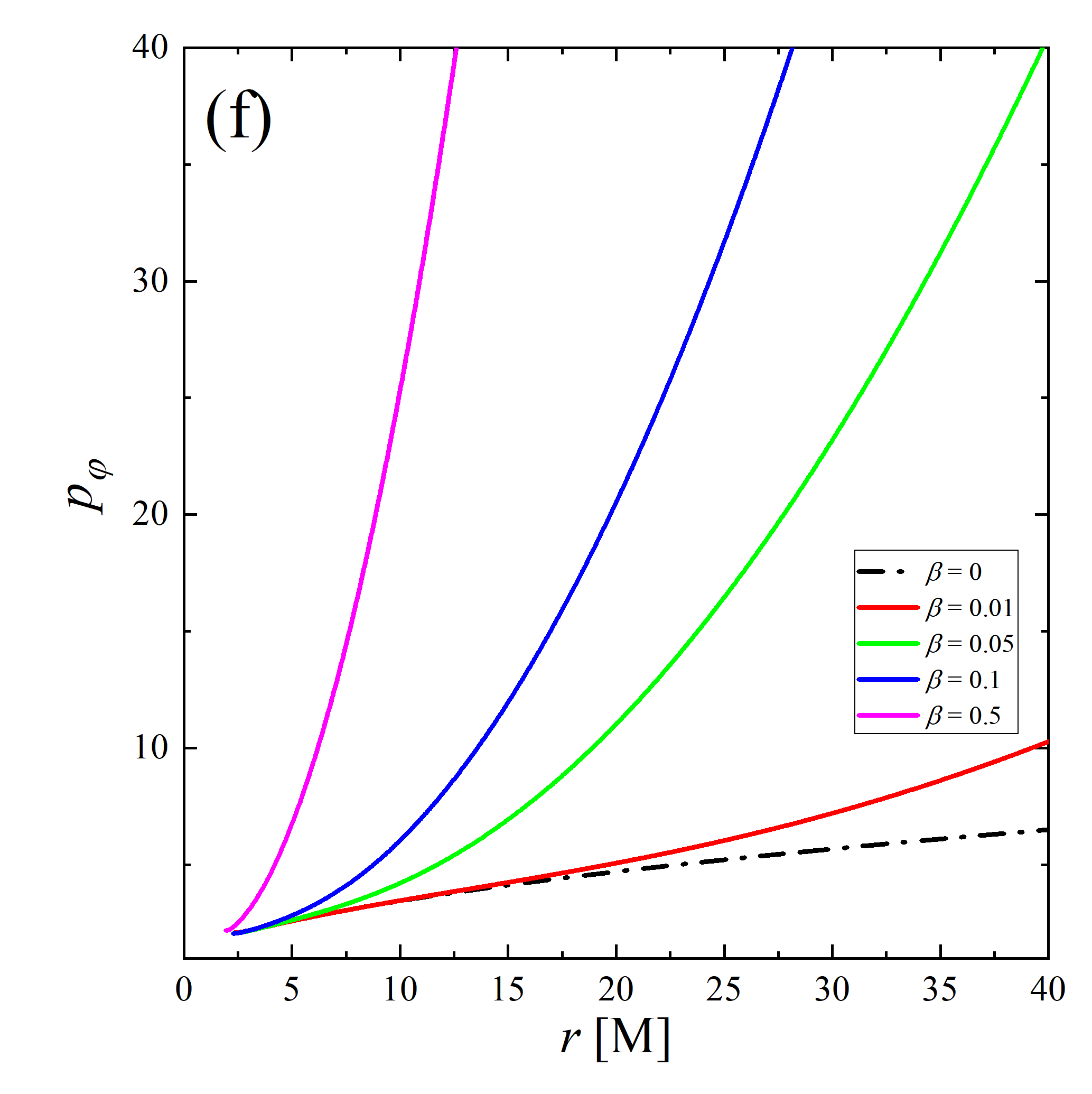}
\includegraphics[width=5cm]{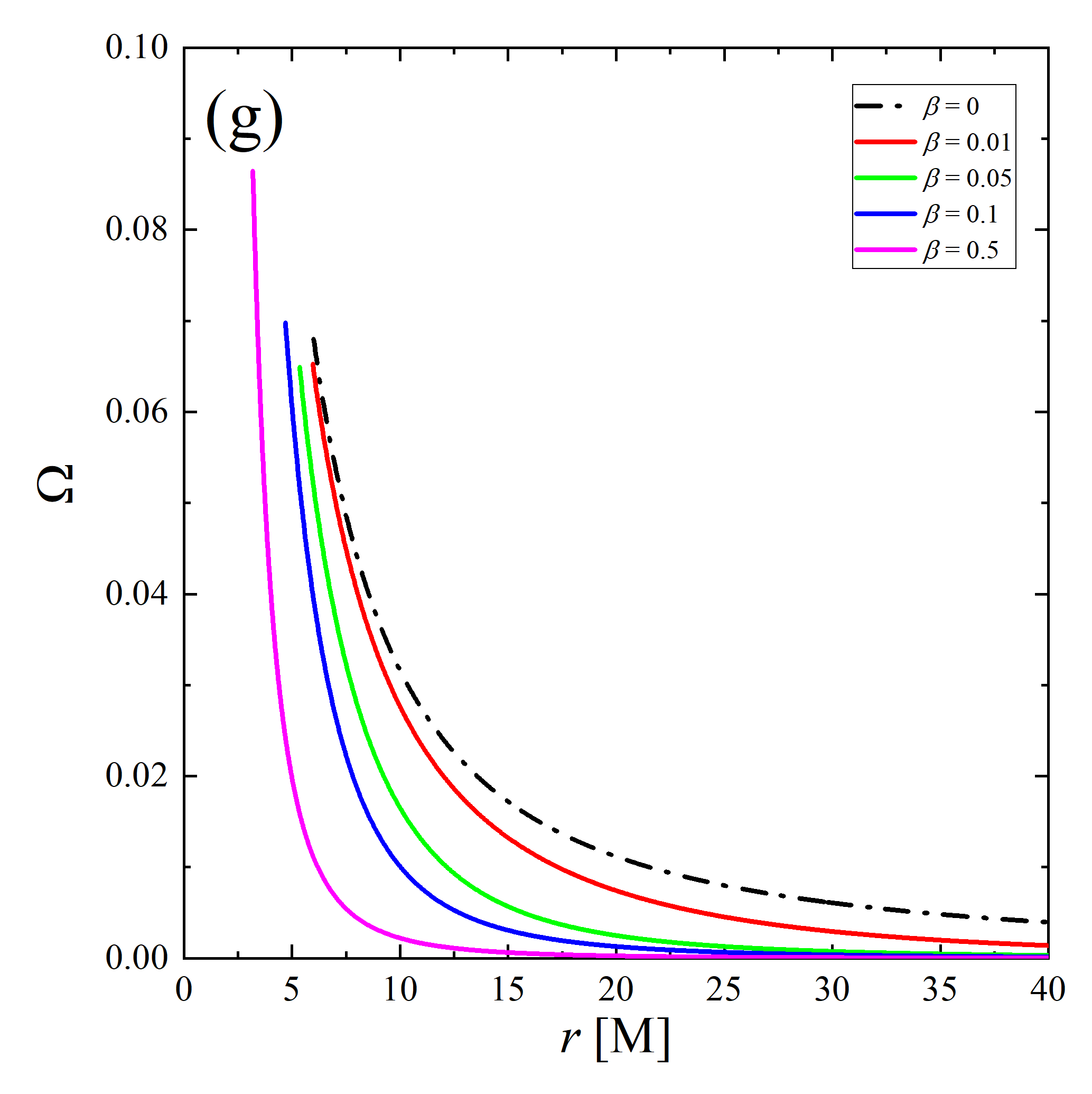}
\includegraphics[width=5cm]{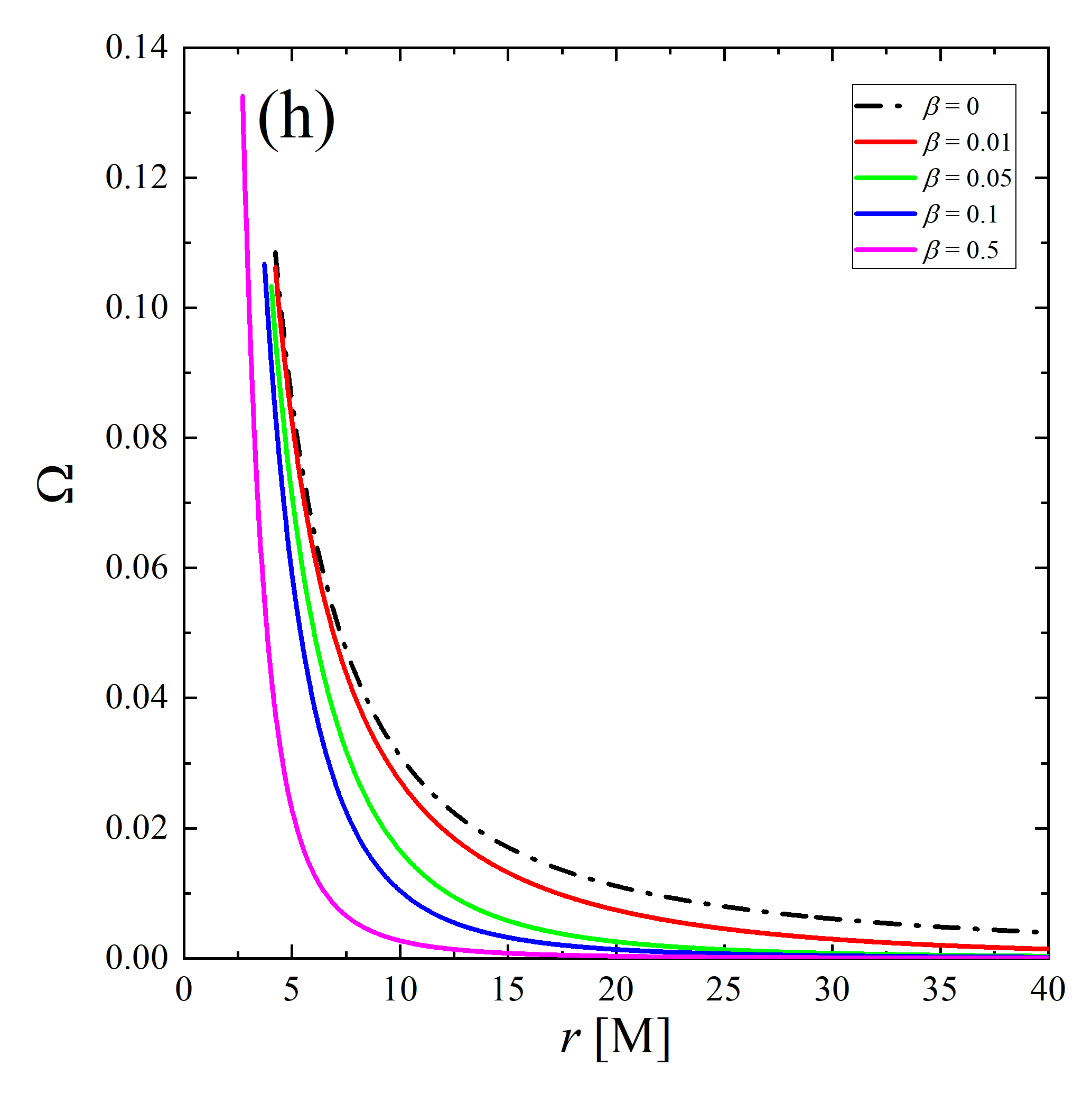}
\includegraphics[width=5cm]{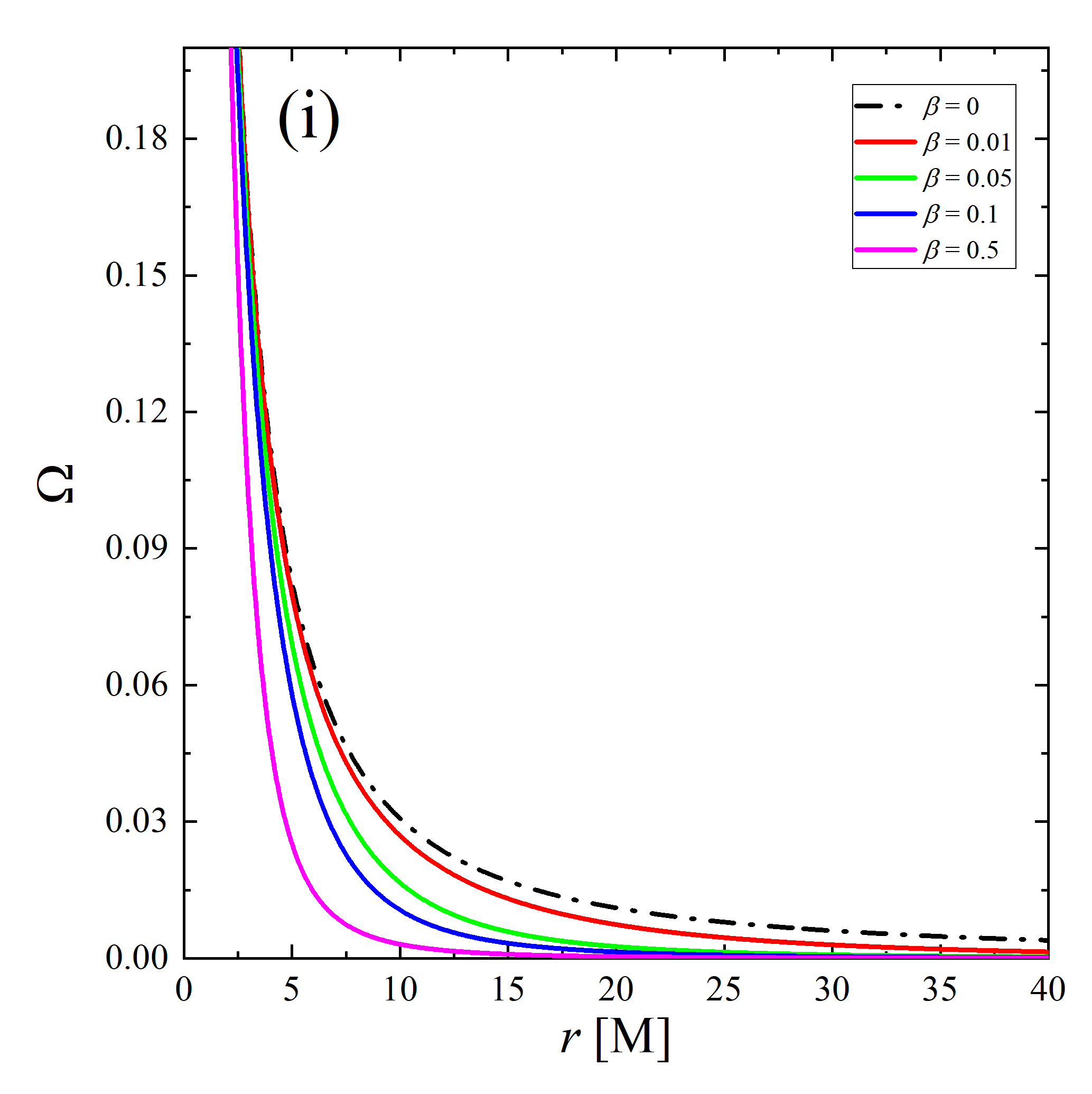}
\caption{(colour online) Dependencies of specific energy $E$, specific angular momentum $p_{\varphi}$, and angular velocity $\Omega$ on the radial coordinate $r$ for timelike particles in quasi-Keplerian orbits across different parameter spaces. From left to right, the spin parameters are $0$, $0.5$, and $0.9$, respectively. The influence of the magnetic parameter $\beta$ on the specific energy depends on both the spin parameter and orbital radius, showing no universal trend.  Nevertheless, increasing $\beta$ clearly enhances the energy contrast between the inner and outer disk regions, suggesting a connection between disk radiation and the magnetic environment. Additionally, larger $\beta$ values increase the specific angular momentum while suppressing the angular velocity.}} \label{fig2}
\end{figure*}
Beyond the ISCO, the quasi-Keplerian circular orbits comprising the accretion disk must satisfy the conditions $\partial V_{\textrm{eff}}/\partial r =0$ and $\partial^{2} V_{\textrm{eff}}/\partial r^{2} > 0$. For integrable spacetimes, these conditions yield analytical expressions for three key orbital parameters: the specific energy $E$, specific angular momentum $p_{\varphi}$, and angular velocity $\Omega = \dot{\varphi}/\dot{t}$. However, when an external magnetic field is introduced to the spacetime, the analytical expressions for these physical quantities no longer exist due to the lack of sufficient integrals of motion. Consequently, one must resort to numerical methods---such as the Newton iteration scheme---to solve the equations. Following this approach, Fig. 2 presents the radial profiles of specific energy, specific angular momentum, and angular velocity for timelike circular orbits under different magnetic field parameters $\beta$. For non-rotating black holes (panel a), the magnetic field exhibits a monotonic influence on $E$---at fixed orbital radius, the specific energy decreases with increasing field strength, while this simple relationship breaks down when spin is introduced (panels b and c), where the magnetic field's effect on $E$ becomes radius-dependent without a universal pattern. In contrast, the magnetic field's impact on $p_{\varphi}$ and $\Omega$ follows clear trends: for fixed radius, $p_{\varphi}$ increases while $\Omega$ decreases with growing $\beta$. On the other hand, the first row demonstrates that stronger magnetic fields amplify the energy difference between inner and outer disk regions, implying that when particles migrate inward through angular momentum loss, substantial energy is radiated away, which suggests positive correlations among disk luminosity, radiative efficiency, and magnetic field strength.
\section{Accretion disk's features}
Within the Novikov-Thorne accretion disk framework, the energy flux density per unit time $F(r)$ [erg $\textrm{s}^{-1}$ $\textrm{cm}^{-2}$], depends on the black hole accretion rate $\dot{M}$, orbital dynamical parameters, and the reduced metric determinant $A$ in the equatorial plane. This relationship is expressed as:
\begin{equation}\label{12}
F(r) = -\frac{\dot{M}}{4\pi\sqrt{-A}}\frac{\Omega^{\prime}}{\left(E-\Omega p_{\varphi}\right)^{2}}\int^{r}_{r_{\textrm{ISCO}}}\left(E-\Omega p_{\varphi}\right)p_{\varphi}^{\prime}\textrm{d}r,
\end{equation}
where $r$ denotes the radial position of the emitting source, and primes indicate derivatives with respect to $r$. Notably, when $p_{\varphi}$ and $\Omega$ have analytical expressions in terms of $r$ (i.e., for integrable spacetimes), obtaining $p_{\varphi}^{\prime}$ and $\Omega^{\prime}$ is straightforward. However, this approach becomes infeasible with the introduction of an external magnetic field. Instead, we numerically compute these derivatives from discrete data in Fig. 2 using finite-difference methods. Furthermore, applying the Stefan-Boltzmann law yields the disk's temperature profile $T$ [K]:
\begin{equation}\label{13}
T(r) = \left[\frac{F(r)}{\sigma}\right]^{\frac{1}{4}},
\end{equation}
where $\sigma$ is the Stefan-Boltzmann constant.

\begin{figure*}
\center{
\includegraphics[width=5cm]{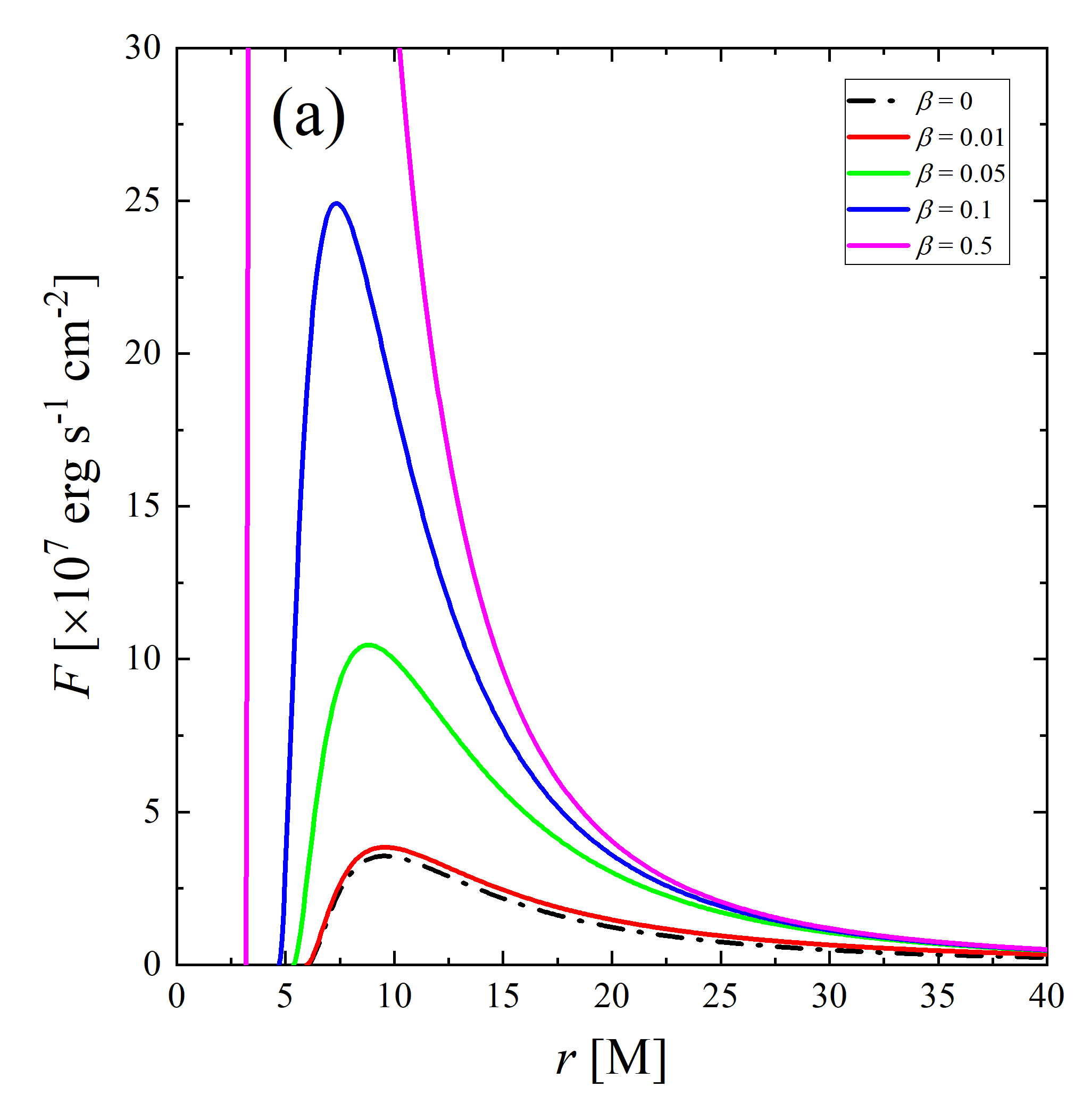}
\includegraphics[width=5cm]{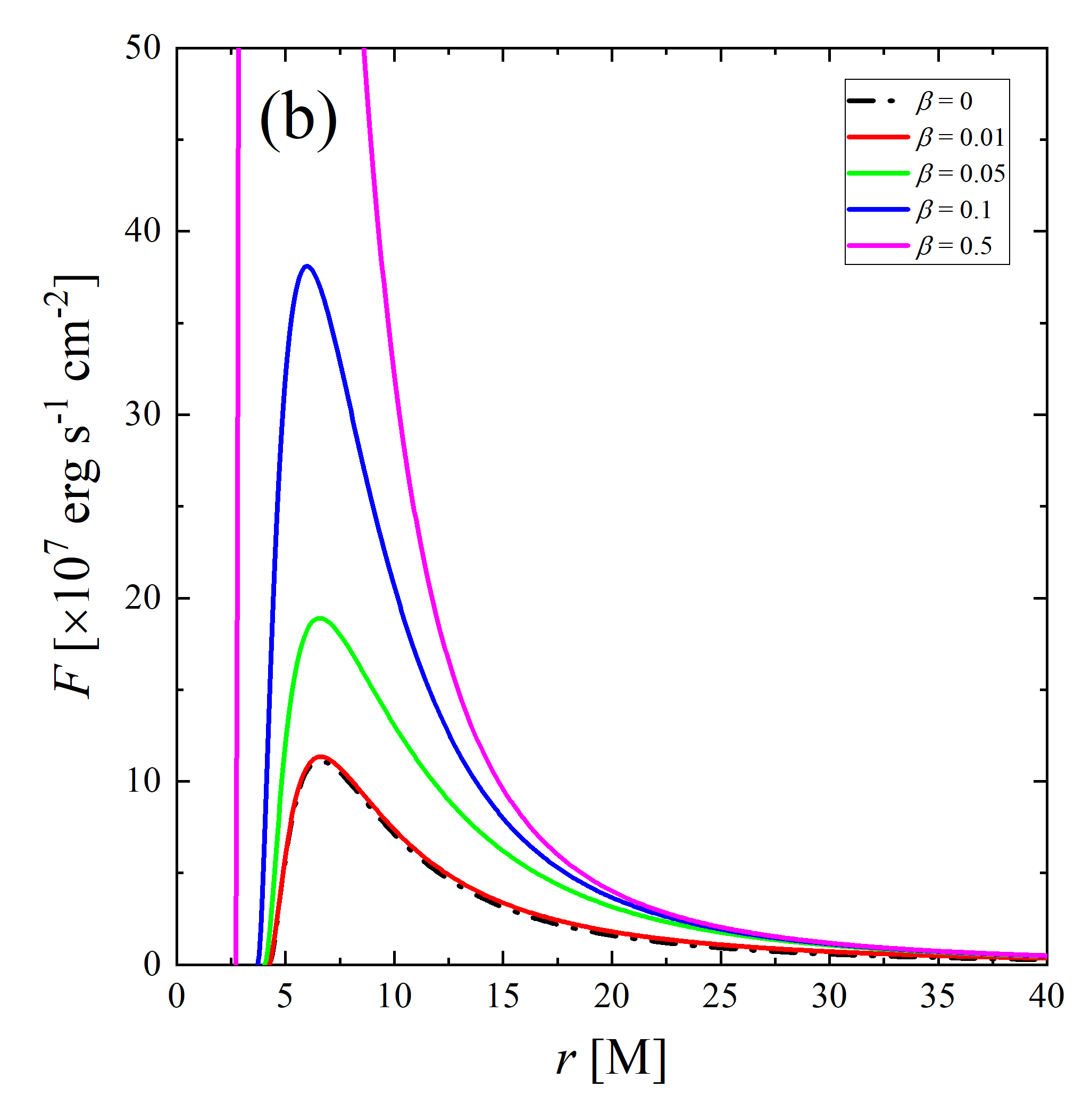}
\includegraphics[width=5cm]{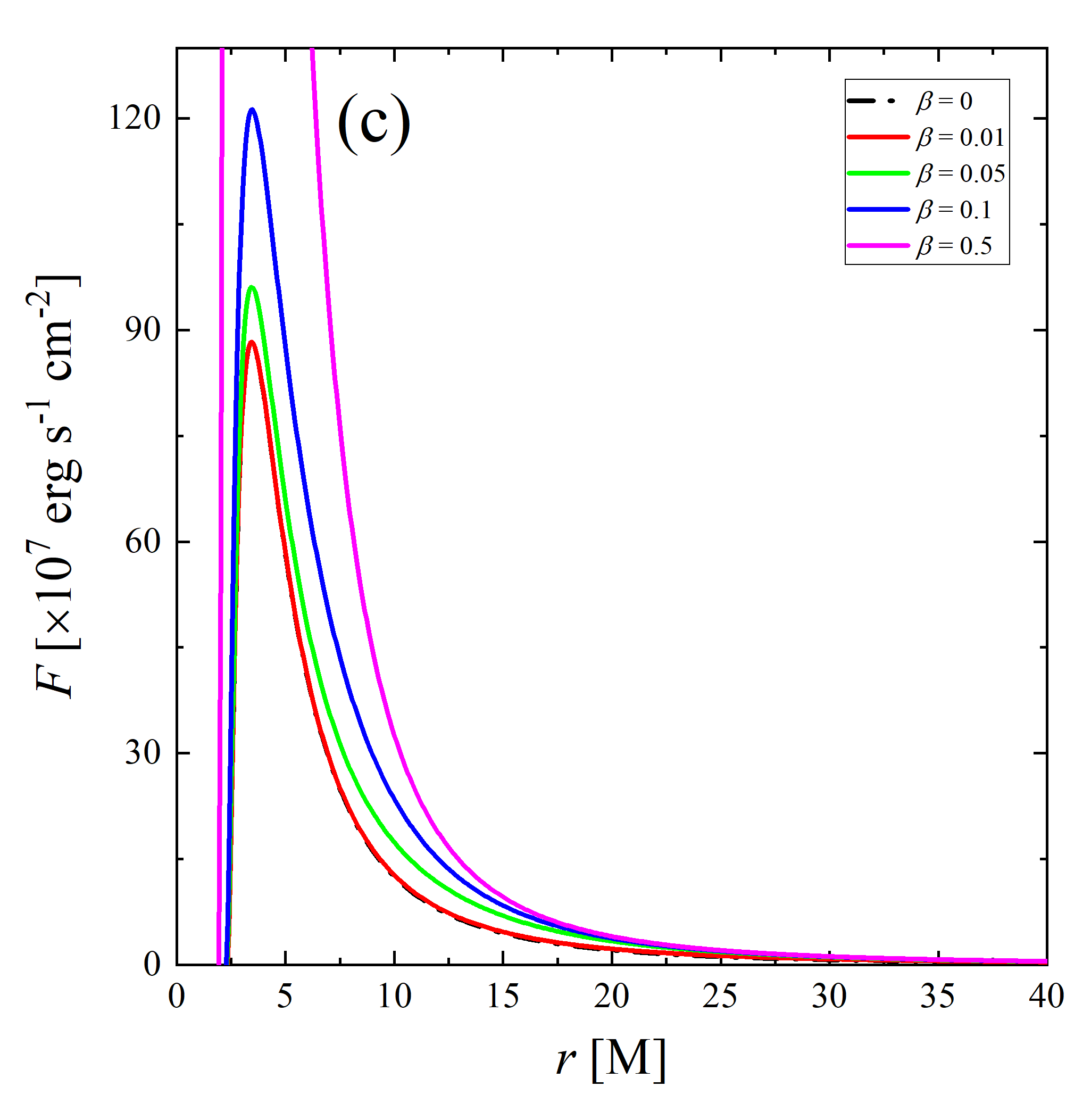}
\includegraphics[width=5cm]{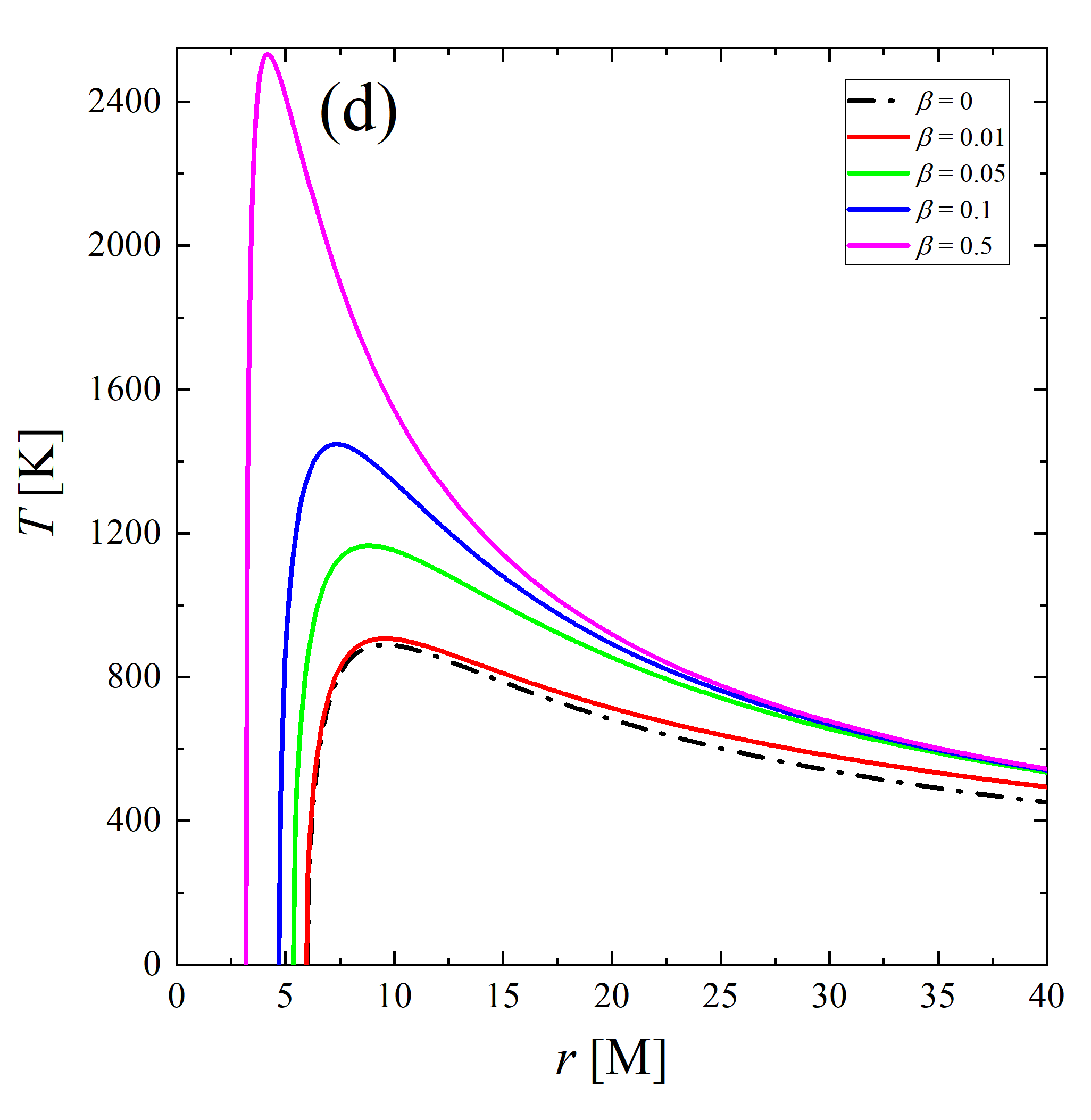}
\includegraphics[width=5cm]{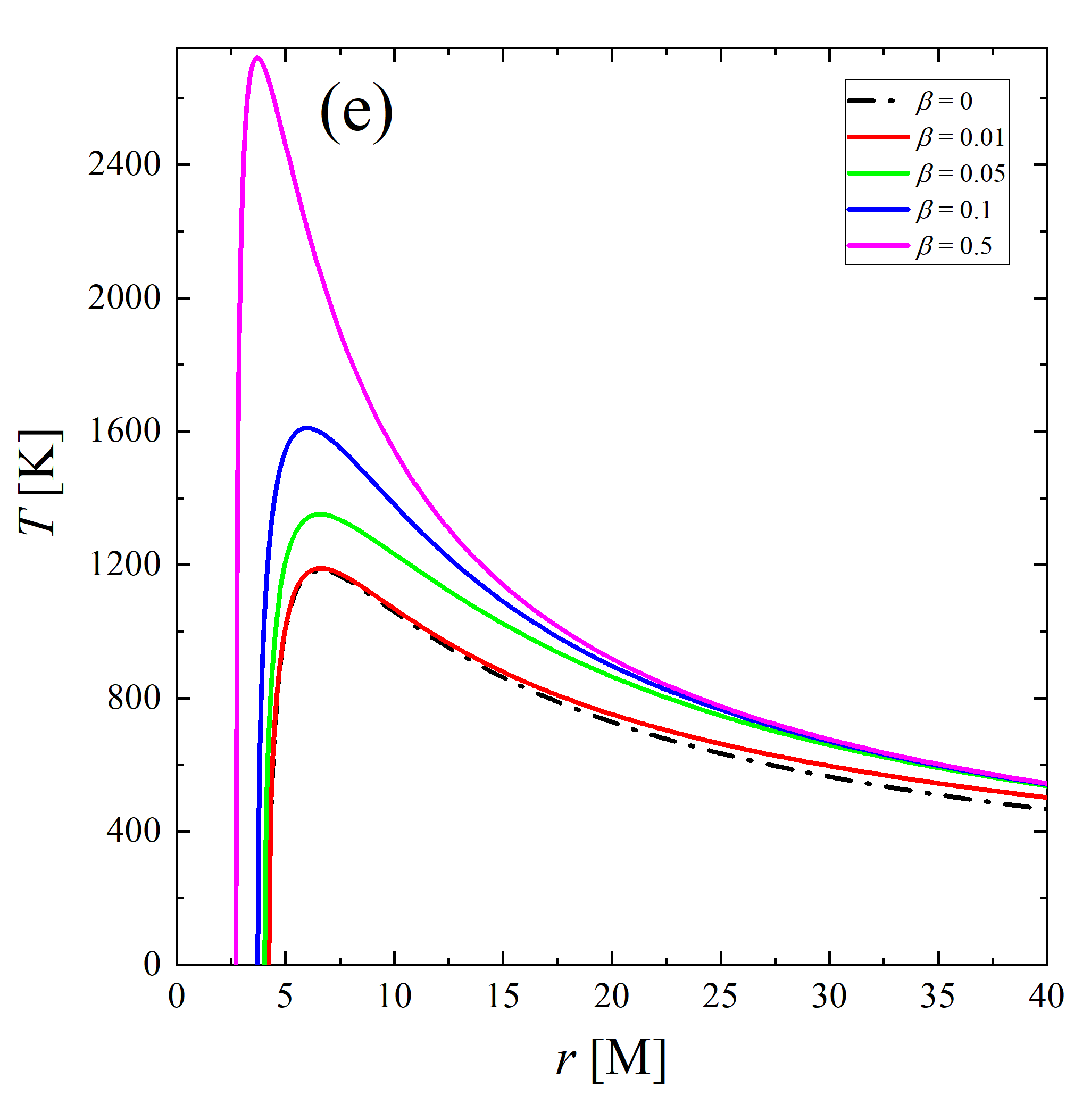}
\includegraphics[width=5cm]{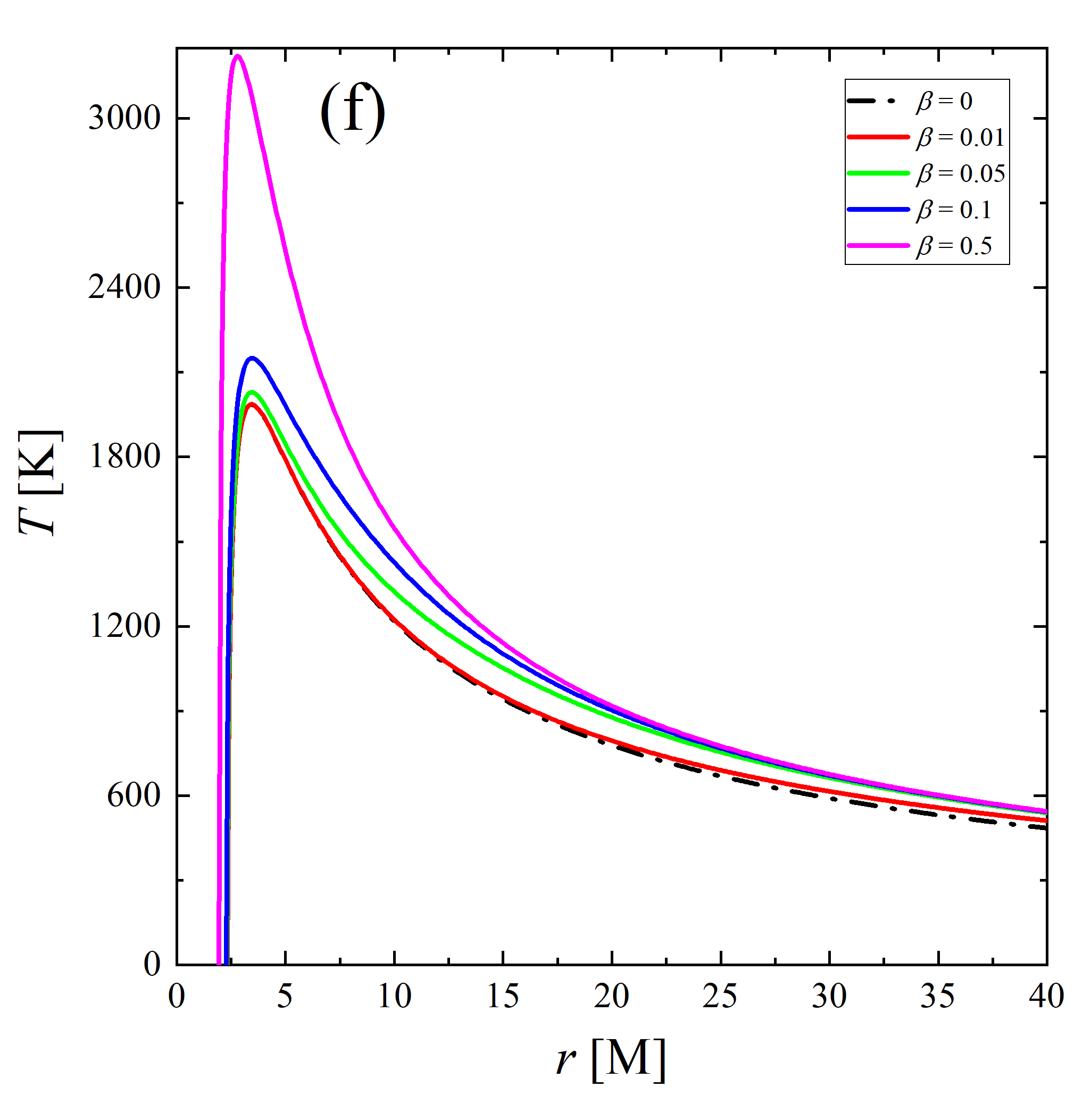}
\caption{(colour online) Radial profiles of the energy flux density (top row) and temperature (bottom row) for Novikov-Thorne accretion disks under varying magnetic field strengths and spin parameters. From left to light, the spin parameters are $0$, $0.5$, and $0.9$, respectively. Both increasing magnetic field strength and spin enhance the disk's energy flux density, with the magnetic effect being particularly significant for low spins. Due to axis scaling, the curves for $\beta=0.5$ (pink) are not fully displayed---their peaks in panels (a)--(c) reach $233$, $310$, and $610$, respectively. The combined influence of spin and magnetic fields on $F$ directly leads to the temperature trends shown in the bottom row: both parameters promote higher disk temperatures.}} \label{fig3}
\end{figure*}
We consider a black hole with mass $M = 10^{6}$$M_{\odot}$ and an accretion rate $\dot{M} = 10^{-12}$$M_{\odot}$$\textrm{yr}^{-1}$, where $M_{\odot}$ represents the solar mass. Our numerical simulations reveal the radial profiles of the disk's energy flux density $F(r)$ and temperature $T(r)$ across different parameter spaces, presented in the first and second rows of Fig. 3, respectively. It is found that increasing the magnetic field parameter $\beta$ enhances the disk's energy radiation while shifting the emission peak toward smaller radii. These effects become more pronounced for lower spin parameters, indicating that rapid black hole rotation can partially suppress the magnetic field's influence on the disk's radiative properties. Moreover, when the magnetic parameter is sufficiently small (e.g., $\beta = 0.01$), the resulting differences in radiation characteristics become observationally indistinguishable. Given the direct algebraic relationship between energy flux and temperature, the temperature profiles exhibit trends similar to those shown in the first row: increasing the magnetic parameter $\beta$ leads to higher disk temperatures. Notably, both the energy flux density and temperature show positive correlations with black hole spin. This implies that for a given black hole mass and accretion rate, there exists an upper limit for $F$ and $T$ in Kerr spacetime, since the spin parameter cannot exceed unity. Any observed values exceeding this theoretical maximum could potentially serve as probes of the magnetic field environment.

Using the blackbody radiation model, the frequency-dependent luminosity of the accretion disk can be expressed as:
\begin{equation}\label{14}
L = \frac{8\pi h\cos\gamma}{c^{2}}\int^{r_{\textrm{edge}}}_{r_{\textrm{ISCO}}}\int_{0}^{2\pi}\frac{\nu_{\textrm{e}}^{3}\sqrt{-A}\textrm{d}r\textrm{d}\varphi}{e^{h\nu_{\textrm{e}}/\kappa T(r)}-1}.
\end{equation}
Here, $\gamma$ is the observer's inclination angle (between the line of sight and the black hole's spin axis), $h$ is Planck's constant, $r_{\textrm{edge}}$ denotes the outer boundary of the accretion disk, $\kappa$ is the Boltzmann constant, and $\nu_{\textrm{e}}$ represents the emitted frequency related to the observed frequency $\nu$ via $\nu_{\textrm{e}} = g\nu$, with $g$ being the redshift factor:
\begin{equation}\label{15}
g = \frac{1+\Omega r\sin\varphi\sin\gamma}{\sqrt{-g_{tt}-2\Omega g_{t\varphi}-g_{\varphi\varphi}\Omega^{2}}}.
\end{equation}

\begin{figure*}
\center{
\includegraphics[width=5cm]{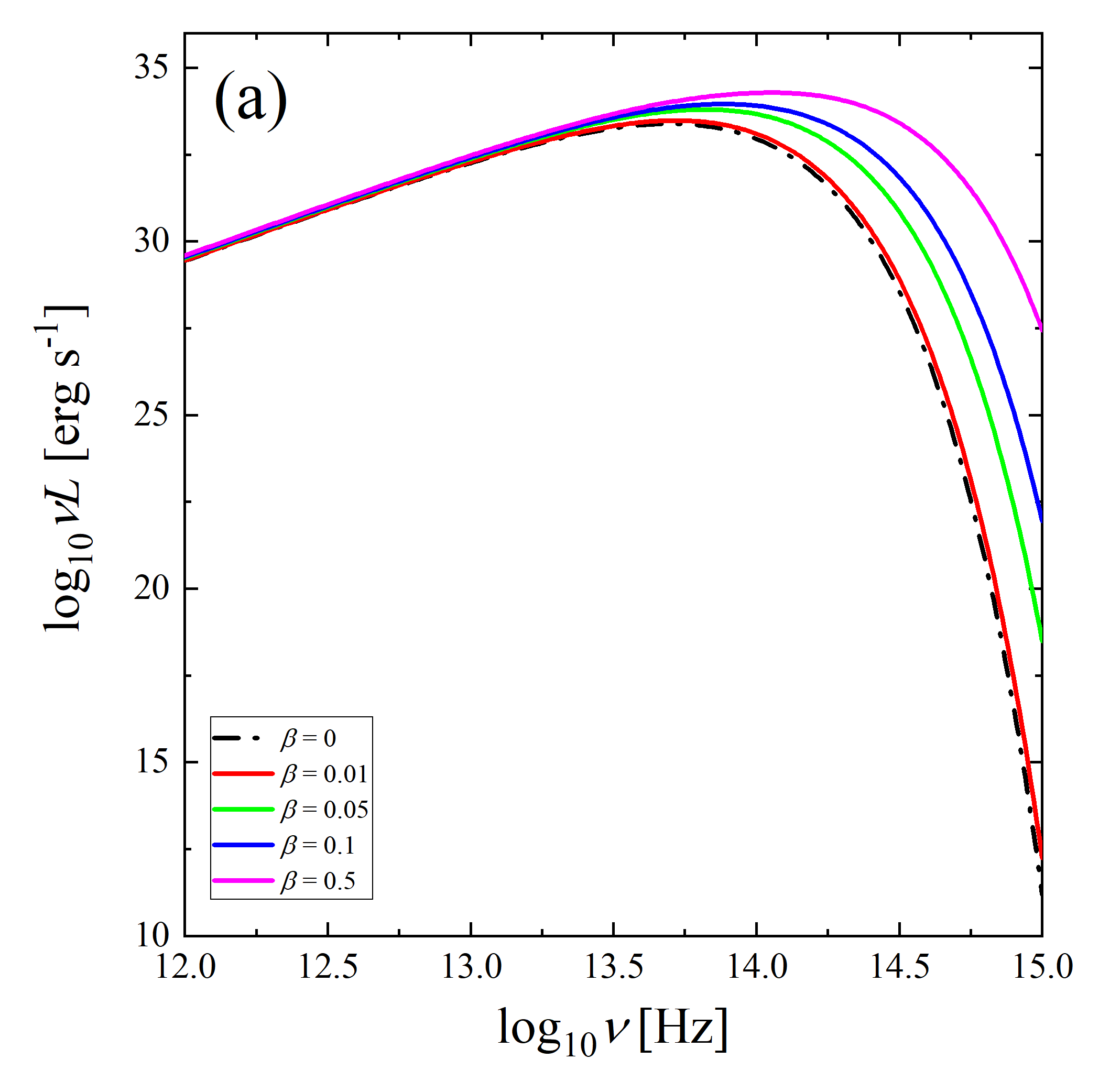}
\includegraphics[width=5cm]{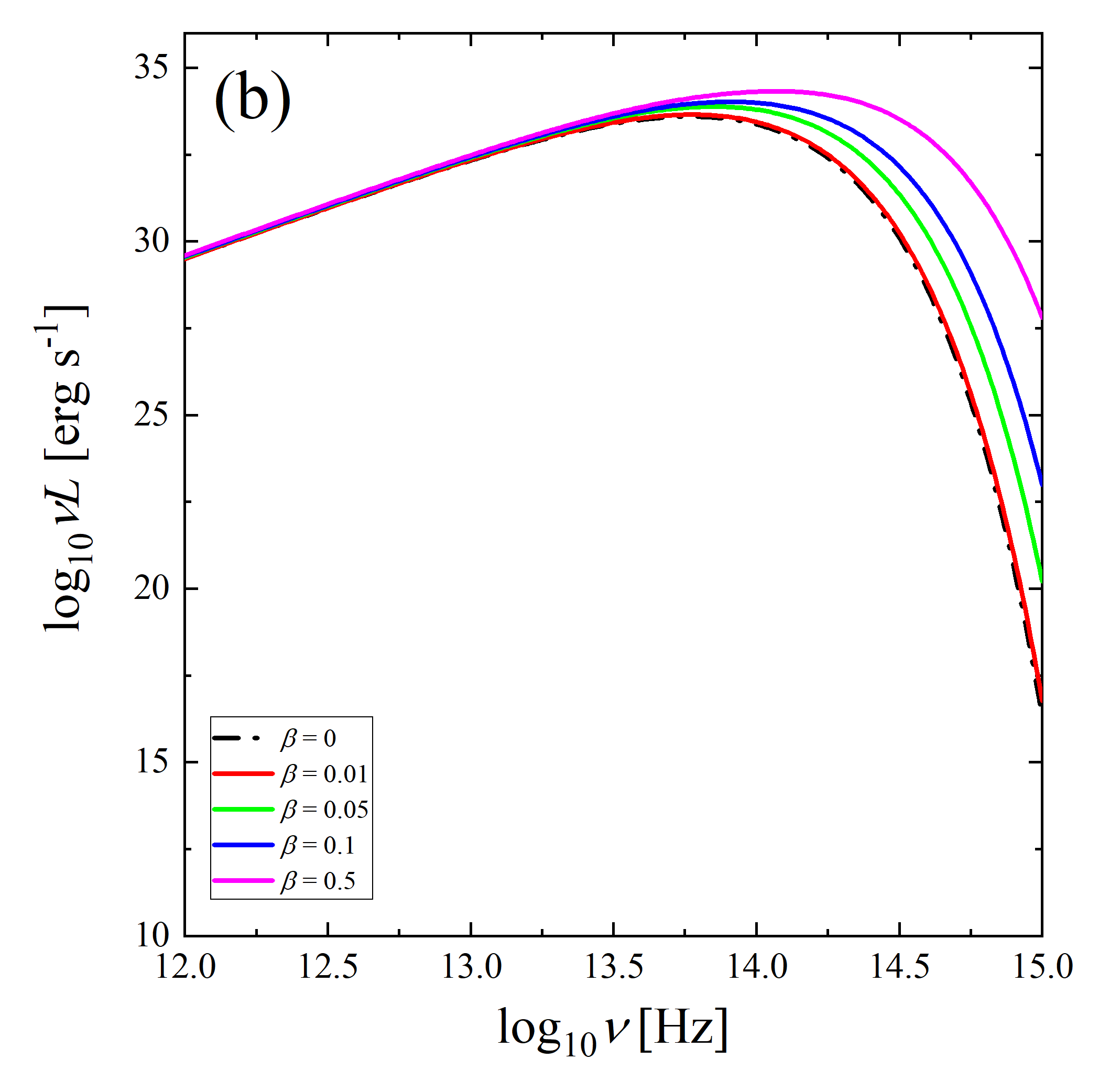}
\includegraphics[width=5cm]{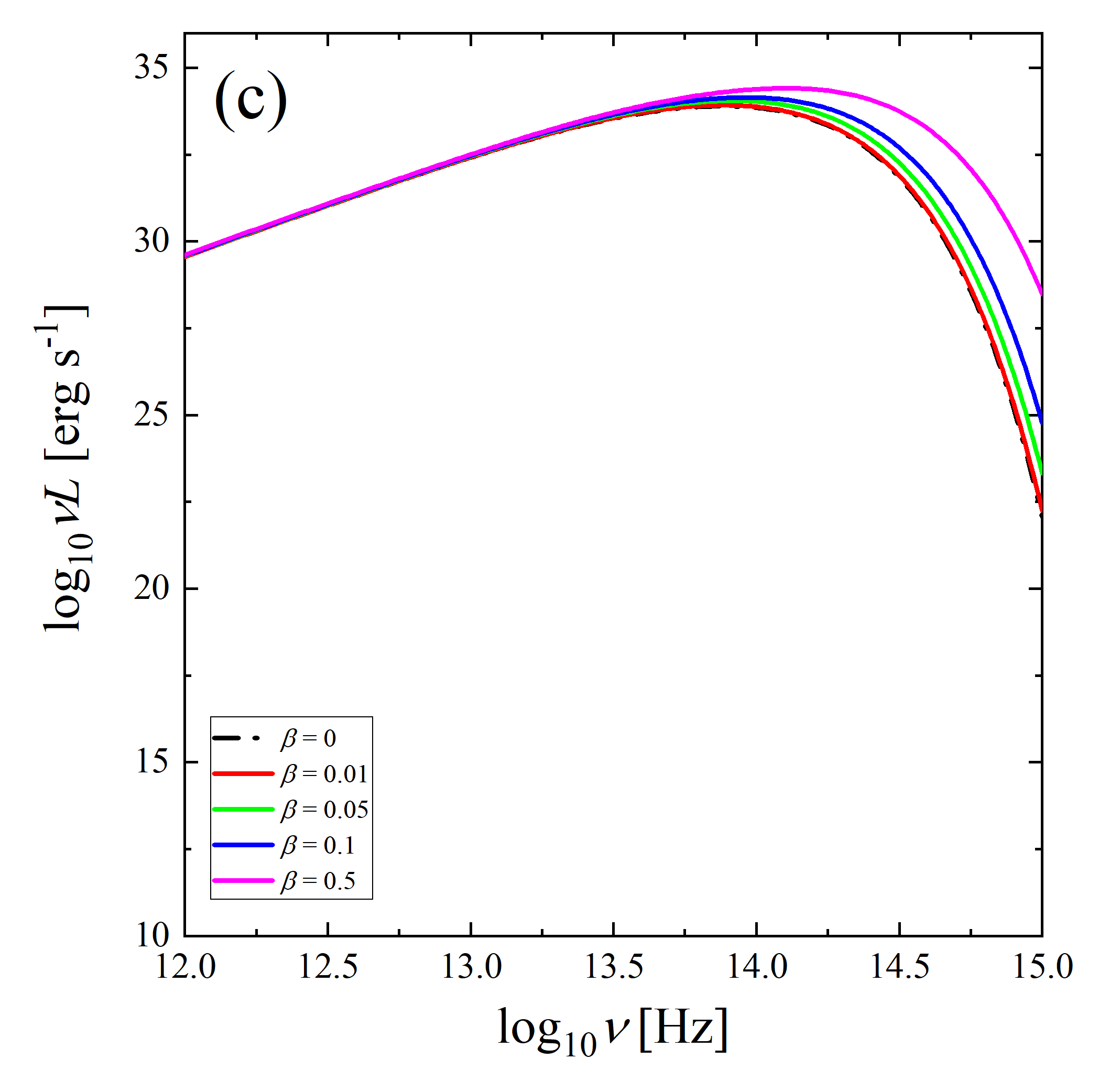}
\includegraphics[width=5cm]{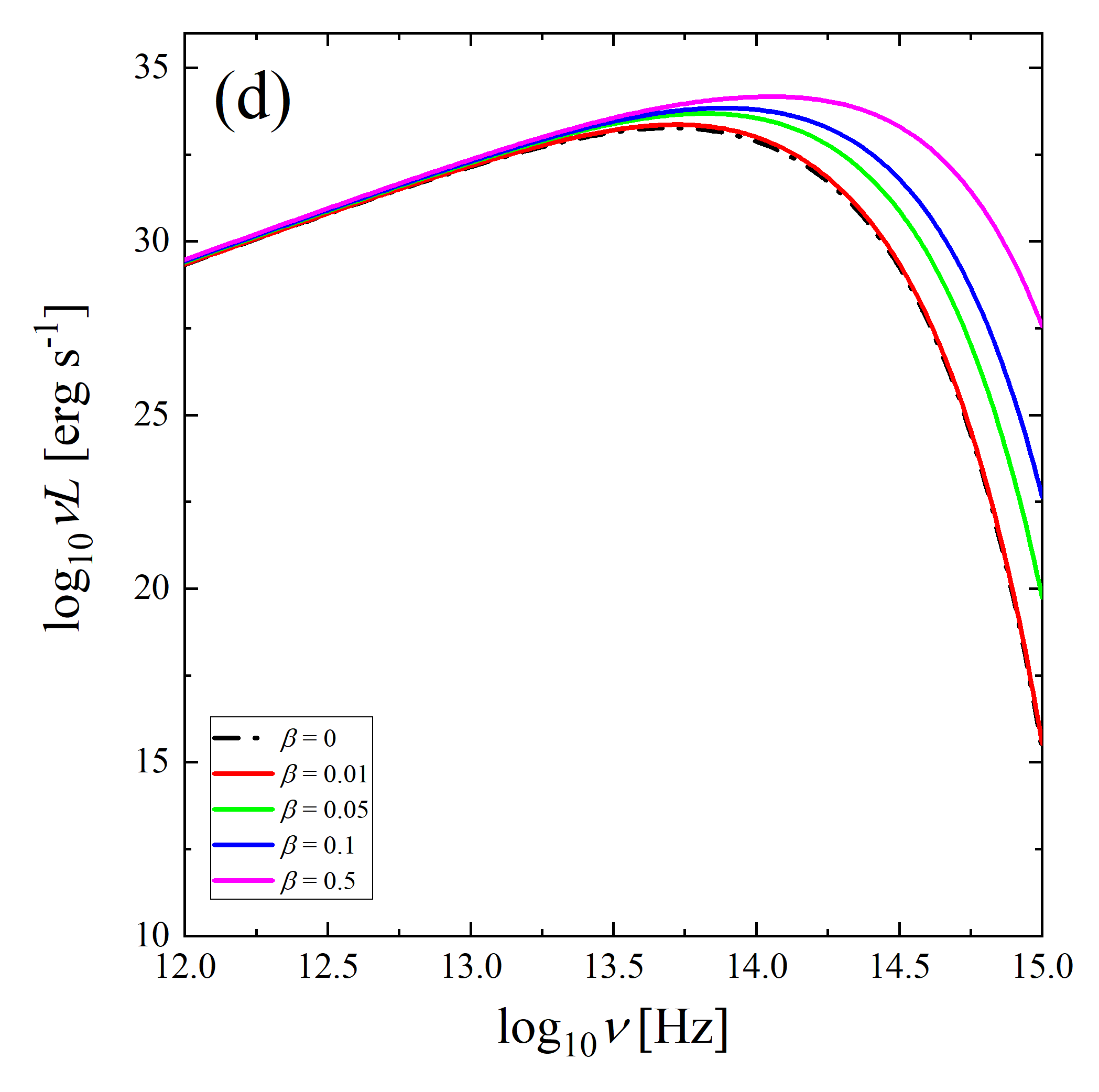}
\includegraphics[width=5cm]{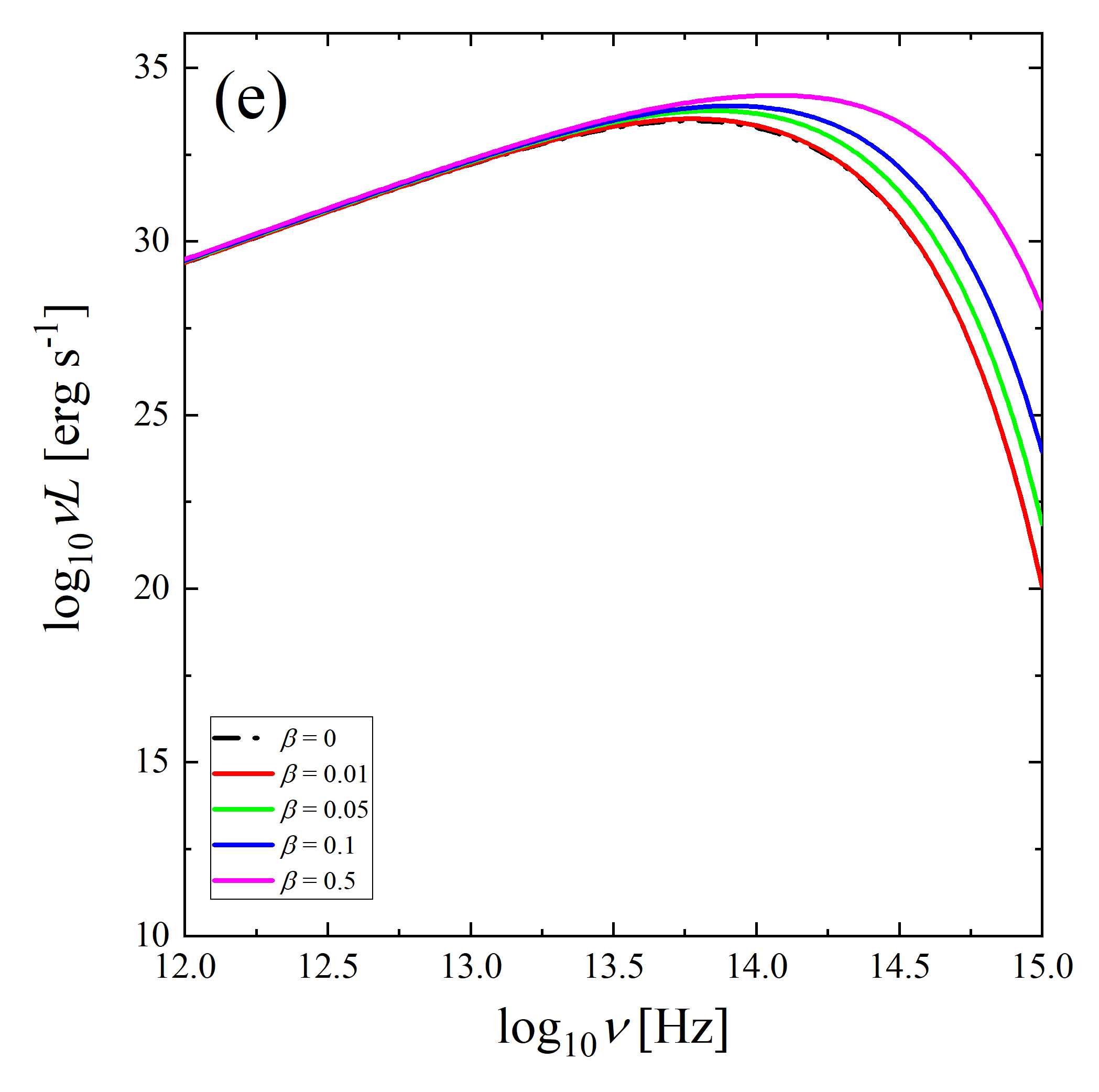}
\includegraphics[width=5cm]{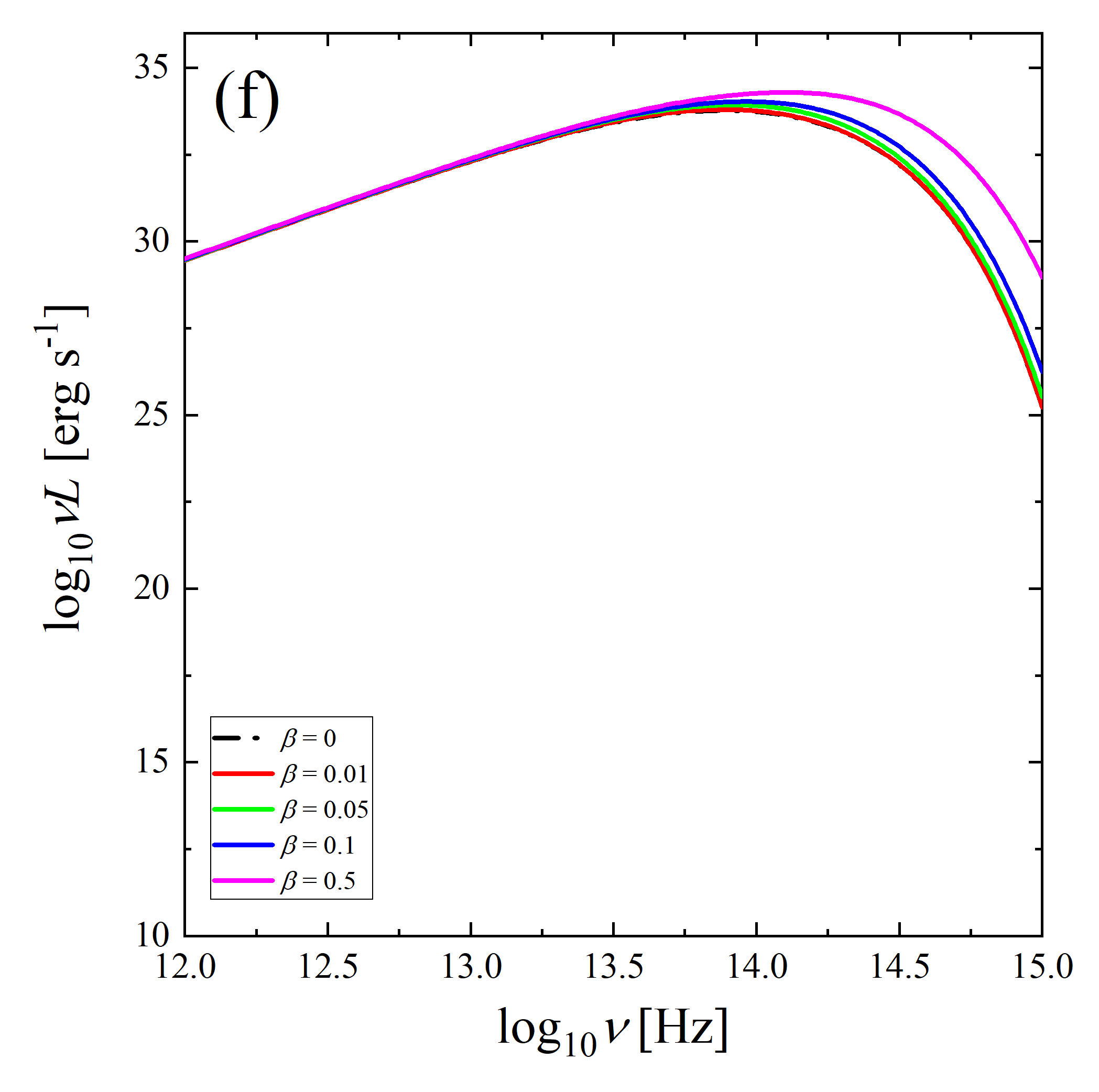}
\includegraphics[width=5cm]{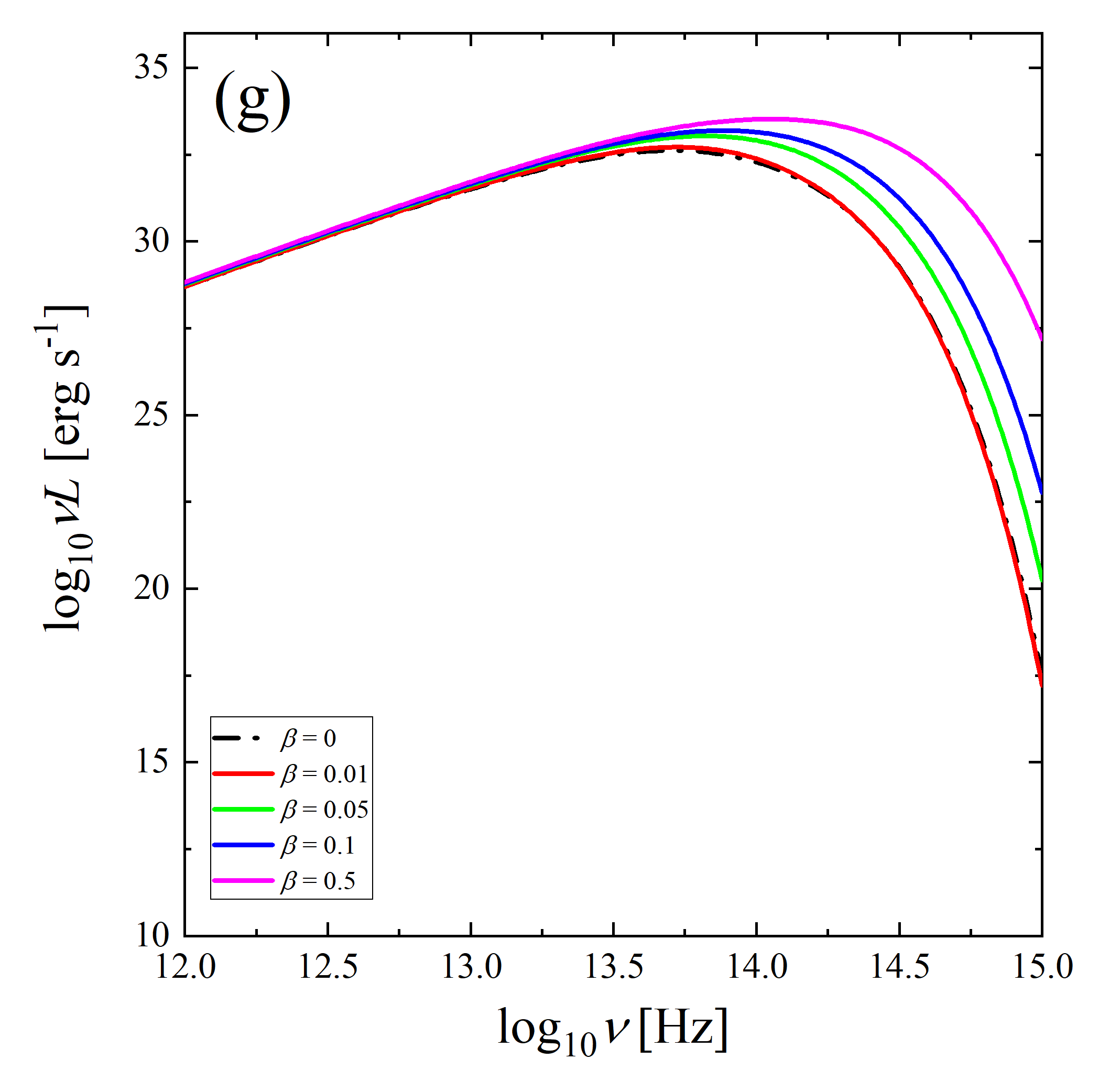}
\includegraphics[width=5cm]{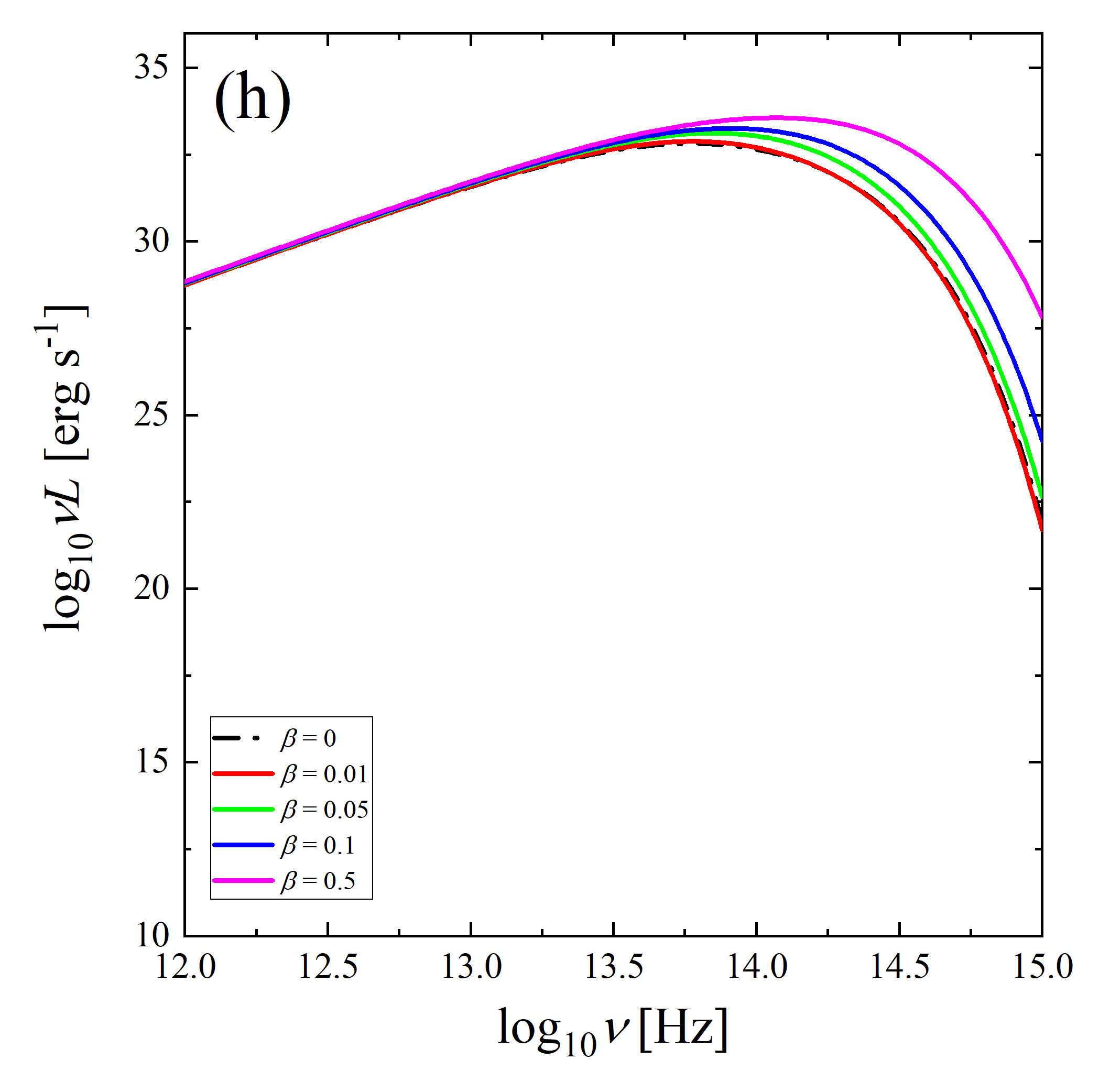}
\includegraphics[width=5cm]{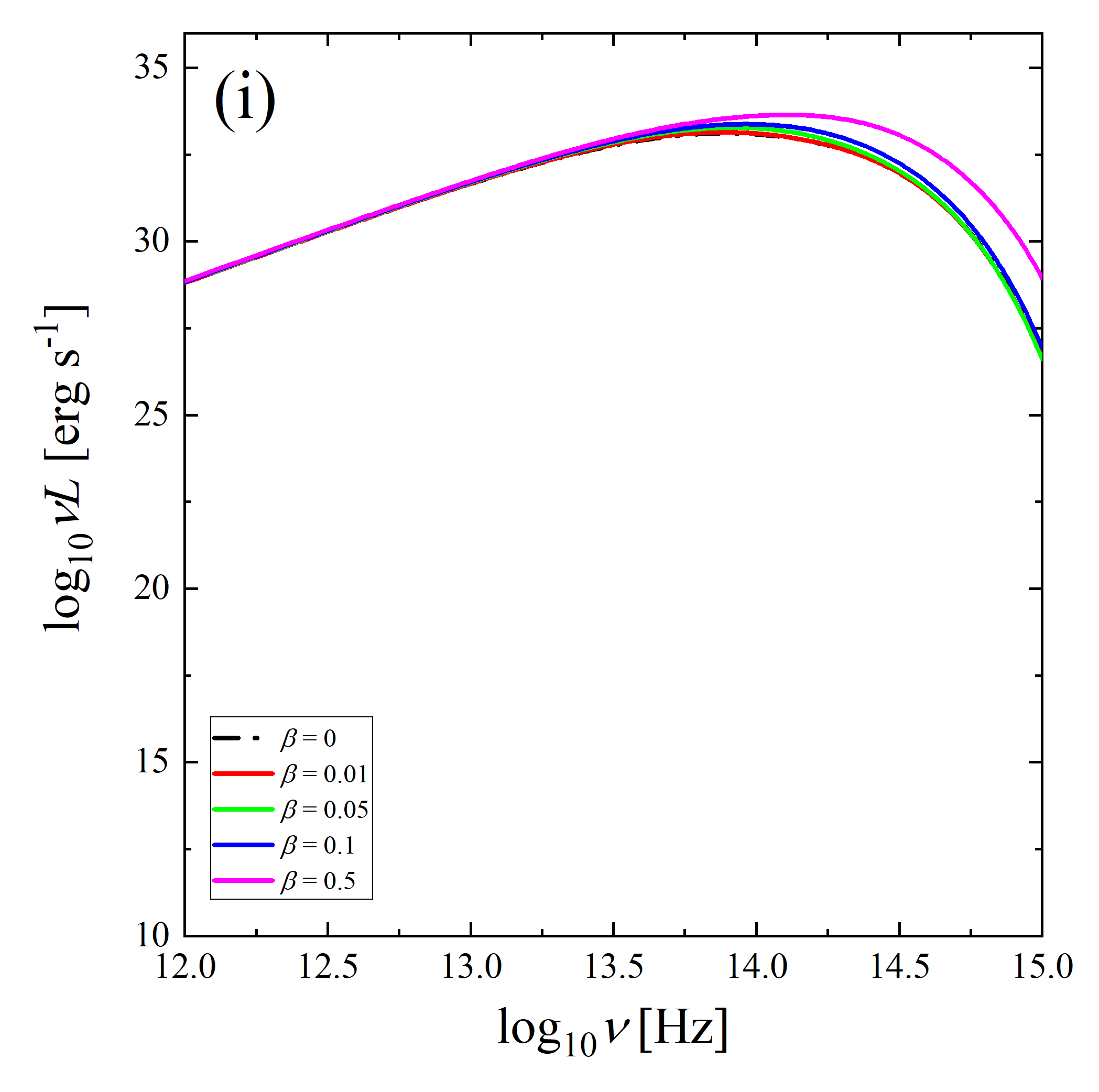}
\caption{(colour online) Dependency of disk luminosity on observational frequency $\nu$ under varying spin parameters ($a = 0, 0.5, 0.9$ from left to right), magnetic parameters $\beta$, and viewing angles ($\gamma = 0^{\circ}, 40^{\circ}$, and $80^{\circ}$ from top to bottom). For the adopted black hole mass and accretion rate, the luminosity peaks in the near-infrared band. Increasing magnetic field strength boosts the spectral peak luminosity and shifts it toward higher frequencies---behavior analogous to spin effects, indicating potential degeneracy between magnetic environment and spin. Notably, spectral variations at $\beta = 0.5$ are distinctly identifiable in most parameter regimes.}} \label{fig4}
\end{figure*}
Adopting the same black hole mass and accretion rate as in Fig. 3, we numerically simulated the blackbody spectra of the accretion disk for varying magnetic field parameters, spin parameters, and observer inclination angles, as presented in Fig. 4. First, it is found that the spectral luminosity of the accretion disk initially rises and then declines with frequency, peaking predominantly within the infrared band ($10^{14}-10^{15}$ Hz), as expected for our adopted black hole mass and accretion rate that yield characteristic disk temperatures of $\sim 10^{3}$ K. Both the peak luminosity and its frequency would shift if different mass or accretion rate were chosen. Then, increasing the observer's inclination reduces the low-frequency luminosity while enhancing the high-frequency component---an effect more pronounced for lower spin parameters. Furthermore, higher black hole spins shift the spectral peak slightly toward higher frequencies and energies, consistent with Wien's displacement law, as increased spin elevates disk temperatures (see Fig. 3, second row). Most notably, stronger magnetic fields significantly boost the total luminosity and shift the peak toward higher frequencies, with particularly dramatic enhancements in the high-frequency regime. However, for rapidly spinning black holes (right column of Fig. 4), magnetic effects---especially those from weaker fields ($\beta \leq 0.1$)---become negligible. We also identify a degeneracy between the impacts of magnetic field strength and spin parameter on the disk's spectral luminosity.

It is crucial to note that for a Kerr black hole with given mass and accretion rate, the peak luminosity of its Novikov-Thorne accretion disk has a theoretical maximum, achieved under maximal spin ($a \rightarrow 1$) with face-on observation ($\gamma = 0^{\circ}$). Any observed luminosity exceeding this limit would strongly indicate the presence of an ambient magnetic field. Furthermore, Fig. 4 reveals that across all parameter spaces, the spectral curve for $\beta = 0.5$ exhibits non-negligible deviations from other cases. This demonstrates that accretion disk luminosity observations can reliably detect magnetic field fluctuations with $\Delta\beta \geq 0.5$. Through the conversion between geometric and SI units \cite{Kopacek and Karas (2014)}, the magnetic field strength is given by
\begin{equation}\label{16}
B_{\textrm{SI}} = \frac{\beta c_{\textrm{SI}}}{\varrho_{\textrm{SI}}\left(\frac{M}{M_{\odot}}\right)1472 \textrm{m}},
\end{equation}
where $\varrho_{\textrm{SI}}$ represents the charge-to-mass ratio of particles. Assuming the accretion disk consists of protons with $\varrho_{\textrm{SI}}=9.5788 \times 10^{7}$ C/kg, we calculate that for a black hole of mass $M = 10^{6}$$M_{\odot}$ and accretion rate $\dot{M} = 10^{-12}$$M_{\odot}$$\textrm{yr}^{-1}$, the case $\beta = 0.5$ corresponds to $B_{\textrm{SI}} = 1.0638 \times 10^{-9}$ T. Under our adopted parameters, this value represents the theoretically detectable threshold magnetic field strength.
\section{CONCLUSIONS AND DISCUSSIONS}
In this work, we numerically determine the specific energy $E$, specific angular momentum $p_{\varphi}$, angular velocity $\Omega$, and particularly the radial derivatives of both $p_{\varphi}$ and $\Omega$ within the Novikov-Thorne accretion disk framework using Newton iteration, finite-difference methods, and interpolation techniques. Our simulations successfully model the energy flux density, temperature, and luminosity of accretion disks in Kerr spacetime with the external electromagnetic field. Most significantly, we establish for the first time the quantitative relationship between magnetic field environments and accretion disk properties in curved spacetime.

The study demonstrates that increasing magnetic field strength enhances the accretion disk's energy flux density and temperature, thereby positively affecting its luminosity. The degree of enhancement depends on photon frequency, spin parameter, and observer inclination angle. This enhancement primarily originates from the inward shift of the disk's inner boundary (the ISCO radius) with stronger magnetic fields. Furthermore, based on how inclination and spin parameters affect disk radiation, we confirm that Kerr spacetime imposes a theoretical maximum on accretion disk luminosity once the mass and accretion rate are determined. Crucially, the presence of magnetic fields can exceed this limit. Therefore, any observed luminosity surpassing the Kerr maximum would provide compelling evidence for the existence of ambient magnetic fields. Moreover, within specific parameter regimes---such as for a black hole with mass $M = 10^{6}$$M_{\odot}$ and accretion rate $\dot{M} = 10^{-12}$$M_{\odot}$$\textrm{yr}^{-1}$---our results demonstrate that accretion disk spectral luminosity observations can probe magnetic field strengths down to $1.0638 \times 10^{-9}$ T.

It is crucial to emphasize that our conclusions are derived under the specific configuration where the magnetic field orientation is aligned with the black hole's angular momentum. When the two are misaligned, the resulting modifications to accretion disk radiation present a compelling open question for future research. Furthermore, the identified degeneracy between the spin parameter and magnetic environments necessitates future work to develop observational diagnostics to break this degeneracy.

\acknowledgements
This research has been supported by the National Natural Science Foundation of China [Grant No. 12403081].

\end{document}